\documentclass[a4paper,11pt]{article}
\pdfoutput=1 
\usepackage[latin1]{inputenc}
\usepackage[T1]{fontenc} % if needed
\usepackage{graphicx, multirow}
\usepackage{subcaption}
\usepackage{float}
\usepackage{bm}
\usepackage{braket}
\usepackage{amsmath}
\usepackage{amssymb}
\usepackage{amscd}
\usepackage{booktabs}
\usepackage{latexsym}
\usepackage{slashed}
\usepackage{color, xcolor}
\usepackage{graphicx}
\usepackage[normalem]{ulem}
\definecolor{orange}{cmyk}{0,0.5,1,0}
\usepackage{verbatim}
\usepackage{rotating}
\usepackage{cancel}
\usepackage{caption}
\usepackage{hyperref}
\usepackage{lineno}
\usepackage{makecell,cite}
\hypersetup{
   colorlinks=true,     % false: boxed links; true: colored links
   linkcolor=blue,      % color of internal links
   citecolor=red,       % color of links to bibliography
   filecolor=magenta,   % color of file links
   }
\newcommand{\beq}{\begin{equation}}
\newcommand{\eeq}{\end{equation}}
\newcommand{\bea}{\begin{eqnarray}}
\newcommand{\eea}{\end{eqnarray}}

\usepackage{tikz}
\usetikzlibrary{arrows.meta, positioning, fit, backgrounds}

\tikzset{
    >=Latex,
    font=\sffamily,
    block/.style={
        rectangle, rounded corners=3pt, draw=black!70,
        fill=gray!8, very thick,
        text width=4.2cm, align=center,
        minimum height=1.6cm
    },
    branch/.style={
        rectangle, rounded corners=3pt, draw=black!60,
        fill=blue!5, thick,
        text width=4.5cm, align=center,
        minimum height=2.2cm
    },
    kinbranch/.style={
        rectangle, rounded corners=3pt, draw=black!60,
        fill=green!6, thick,
        text width=4.2cm, align=center,
        minimum height=1.8cm
    },
    fusion/.style={
        rectangle, rounded corners=3pt, draw=black!80,
        fill=orange!10, very thick,
        text width=5cm, align=center,
        minimum height=2.5cm
    },
    output/.style={
        rectangle, rounded corners=3pt, draw=black!90,
        fill=red!8, very thick,
        text width=3.5cm, align=center,
        minimum height=1.5cm
    },
    arrow/.style={->, thick}
}
\begin{document}
%\title{}  
%\author{}
%\affiliation{}
%\author{}
%\affiliation{} 
%\author{}
%\affiliation{} 
%---------------------------------------
\begin{center}

%{\large \bf Probing Heavy Resonances Through Higgs Boson Pair Production in Vector Boson Fusion} \\
{\large \bf Machine Learning Enhanced Detection \\[0.15cm]
of Higgs Chain Decays in Vector Boson Fusion} \\ 
\vskip 0.6cm
Shreecheta Chowdhury$^{a}$\footnote{shreecheta$_{-}$c@srmap.edu.in},
Amit Chakraborty$^{a}$\footnote{amit.c@srmap.edu.in},
and
Stefano Moretti$^{b,}$$^{c}$\footnote{stefano.moretti@cern.ch}
\vskip 0.6cm
{$^a$Department of Physics, SRM University-AP, Amaravati, Mangalagiri 522240, India}
\vskip 0.1cm
{$^b$School of Physics \& Astronomy, University of Southampton, Southampton SO17 1BJ, UK } 
\vskip 0.1cm
{$^c$Department of Physics \& Astronomy, Uppsala University, Box 516, 751 20 Uppsala, Sweden}
\end{center}
\vskip 0.3cm

\begin{abstract}
Over the years, Vector Boson Fusion (VBF) has established itself as one of the most robust production channels for studying the Higgs boson, while also serving as a promising pathway for exploring potential signatures of physics Beyond the Standard Model (BSM) at the Large Hadron Collider (LHC). Following the discovery of a SM-like Higgs boson, new opportunities have arisen to also  investigate heavy resonances that decay into SM-like Higgs boson pairs, $hh$, thereby offering valuable insights into the structure of the Higgs sector and the dynamics governing Electro-Weak Symmetry Breaking (EWSB). In this work, we analyze a final state involving, alongside 2 forward/backward light quarks, 4 $b$-quarks emerging from the chain decay $h_2\to h_1h_1\to b\bar b b\bar b$ wherein the heavy CP-even Higgs state $h_2$  is produced in the VBF process $qq\to qqh_2$ and belongs to the Next-to-Minimal Supersymmetric SM (NMSSM). This BSM scenario is used as an illustrative example of the potential of using only low-level calorimeter information enhanced by advanced Deep Learning (DL) methodologies in searching for this channel, which can achieve a statistical significance of approximately $4.5\sigma$, for an integrated luminosity of 300 fb$^{-1}$ at the CERN machine.
\end{abstract}

\newpage
%\vskip 1.5cm
%%%%%%%%%%%%%%%%%%%%%% Table of content %%%%%%%
\hrule
\tableofcontents
\vskip 1.0cm
\hrule

%%%%%%%%%%%%%%%%%%%%%%%%%%%%%%%%%%%%%%%%%%%%%%%
\newpage

\section{Introduction}  \label{intro} 
%======================================================

The discovery of the Higgs boson with a mass of approximately 125 GeV \cite{Higgs_discovery,CMS_Higgs} confirmed a fundamental piece of the SM while simultaneously highlighting its limitations. In fact,  the SM shows several  experimental limitations and theoretical inconsistencies, as it fails to explain, e.g., the existence of Dark Matter (DM) and the observed matter-antimatter asymmetry of the Universe as well as the huge discrepancy between the EW and Planck scales (the so-called `hierarchy problem') and the absence of 
coupling unification, respectively. In short, all these problems demand an extension of the SM. In fact, these shortcomings may suggest that the aforementioned SM-like Higgs boson is not just an isolated discovery, but rather a portal to BSM physics. 

In the SM, the Higgs boson couples to fermions (like quarks and leptons) through Yukawa interactions and to gauge bosons ($W$ and $Z$) through the gauge charges. At the LHC, the dominant mode to produce such a Higgs boson is gluon-gluon fusion (ggF), via a virtual loop of heavy top quarks, producing the SM-like  Higgs state ($h_1$) as well as potentially other Higgs states. Furthermore, a numerically comparable production mode for Higgs bosons, albeit limited to the case of CP-even ones, is VBF. If one is interested in studying SM-like Higgs boson pair (hereafter, di-Higgs) production for the purpose of accessing the Higgs trilinear self-couplings, in turn shedding light on the shape of the Higgs potential, both such processes play a crucial role, as they could offer complementary means (the former via Yukawa couplings and the latter via gauge ones) of producing a heavy Higgs state ($h_2$), in turn decaying into two SM-like Higgs states, i.e., $h_2\to h_1h_1$, indeed, involving the Higgs trilinear self-coupling $h_2h_1h_1$.

Looking BSM, a theory which is well motivated is Supersymmetry (SUSY), as it can remedy most, if not all, of the above SM flaws. Already in its minimal realisation,  known as the Minimal Supersymmetric SM (MSSM) \cite{Nilles1984,Lykken1996,WessBagger1992,Chungetal2005,HaberKane1985,Martin1998,Drees2005,BaerTata2006}), it predicts the existence of multiple Higgs boson,
since it requires a Higgs sector made up by two doublet fields. However, the MSSM is strongly constrained by a variety of experimental data, so that it has slowly fallen from favour as a phenomenologically viable incarnation of SUSY. Specifically, its Higgs sector is such that the decay discussed above, $h_2\to h_1h_1$ is no longer possible with any significant rate. 
The Next-to-MSSM (NMSSM), while retaining all the MSSM benefits in terms of
remedying SM flaws, expands the particle spectrum by introducing an additional gauge singlet field, which couples to the physical fields emerging from the two Higgs doublets with a strength $\lambda$, a new free parameter (for reviews of the NMSSM see \cite{Maniatis2010,Ellwanger2010,Moretti:2019ulc}). Crucially, this coupling $\lambda$ provides a direct tree level contribution to the SM-like Higgs mass,  $m_{h_1}$, reducing the reliance on the heavy top squark loops needed in the MSSM to uplift its lowest order prediction for the lightest Higgs boson mass of the MSSM from $m_Z$ to 125 GeV,
thus relieving  this so-called `small hierarchy problem' of the MSSM. In fact, the NMSSM also remove the $\mu$-problem of the MSSM (i.e.,  the
unmotivated need to have the Higgsino mass parameter $\mu$  at the EW scale in order to obtain EWSB, while the SUSY theory would demand it to be either zero or at the Planck mass). 
%Additionally, mixing between the Higgs doublet fields and the singlet field allows for singlet-like Higgs bosons that have very weak couplings to SM particles. This provides an interesting production and decay profile of the SM and BSM particles. 
Therefore, the NMSSM provides a more natural and phenomenologically richer framework than both the SM and MSSM for accommodating the observed Higgs boson, as it offers a 125 GeV Higgs boson with little fine-tuning alongside additional lighter and heavier Higgs bosons, both charged and neutral. Specifically, a very interesting aspect of NMSSM Higgs sector, satisfying all current LHC data, is indeed the possibility of having sizable $h_2\to h_1h_1$ decays, as one can easily have  $m_{h_2}>2m_{h_1}$. 
%the modified Higgs decay channels, including the decay of SM-like Higgs boson into lighter Higgs states. The model also accommodates a natural candidate for dark matter - the lightest supersymmetric particle (LSP) that can be a singlino-like particle satisfying all the available collider constraints. 

In Ref.~\cite{Das:2018fog}, an interesting phenomenological work was carried out within the NMSSM Higgs sector. The study explored the NMSSM parameter space consistent with current LHC measurements and analysed the effect of modified couplings on the Higgs production and decay mechanisms that distinguish the NMSSM from other SUSY frameworks.  One of the key findings of this work was to obtain the region of parameter space of the NMSSM wherein the production of the $h_2$ boson discussed above is primarily  through the VBF process, instead of the ggF one, the latter being usually the dominant mode for single Higgs boson production, whether SM-like or otherwise. The reason for this is the reduced coupling of the heavy Higgs boson $h_2$ with the SM top quarks and, therefore, a smaller ggF cross section. %Other possible modes of this Higgs boson are associated production modes, i.e. $b\bar{b}H$, $t\bar{t}H$, $WH/ZH$, with $H$ being the Higgs boson under consideration. 
In \cite{Das:2018fog}, the authors analysed the prospects of observing such relatively heavier Higgs state via VBF at the High-Luminosity LHC (HL-LHC) \cite{Gianotti:2002xx} through a final state consisting of 4 isolated leptons ($\ell^\pm$) and two forward/backward jets produced through the VBF production process. More specifically, the authors have studied the sub-process $ qq \to qq h_2 \to ZZ jj$, where $h_2$ is the second lightest CP-even Higgs boson of the NMSSM, which decays to a pair of $Z$ bosons, in turn yielding $ZZ\to \ell^+\ell^-\ell^+\ell^-$ ($\ell=e, \mu$). However, as pointed out in the same work,  $h_2$ also decays to a pair of SM Higgs bosons ($h_1h_1$), albeit with a somewhat more modest rate, thereby motivating us to perform a complementary study to \cite{Das:2018fog} by looking at the sensitivity of the $h_2 \to h_1 h_1$ decay mode,
with both of the SM-like Higgs bosons decaying to a pair of $b$-quarks, i.e., the advocated $h_2\to h_1h_1\to b\bar b b\bar b$ signature.  
%%%%%%%%%%%%%%%%%%%%%%%textcolor{violet}upto here%%%%%%%%%%%%%

Here, we plan to investigate the prospects for discovering a heavy Higgs boson $h_2$ of mass 700 GeV or so within the NMSSM framework (essentially, the Benchmark Point (BP) C of Ref.~\cite{Das:2018fog}), produced via VBF and decaying through a pair of SM-like Higgs boson which further decays to a pair of $b$-quarks at Run 3 of the LHC. The significant mass difference between $h_1$ and $h_2$ allowed in the NMSSM (and reflected by the mentioned BP) provides a significant boost to $h_1$ and thus the decay products of each of these Higgs states can form a spray of hadrons, i.e., a boosted $b$-jet. It is therefore reasonable to cluster these events using a large-$R$ jet algorithm and later analyze the energy distribution among these jets using the jet sub-structure technique. A conventional cut-based analysis, which we present  later, yields a statistical significance of merely 1.7$\sigma$, which thus falls well short of any discovery threshold. We then introduce a Machine Learning (ML) (actually, a DL) based algorithm in a search strategy that deploys a Multi-layer Convolutional Neural Networks (CNN) to learn 
the distinctive footprint of boosted di-Higgs decay products through jet images. We consider both fixed radius (fixed-$R$) and variable radius (variable-$R$) clustering algorithms and two separate approaches to reconstruct the large-$R$ jets from the flow of objects into the electromagnetic and hadronic calorimeters. This Multi-layer CNN achieves a significance of 2.8$\sigma$, which is a clear improvement over conventional cut-based approaches. Furthermore, to leverage the energy profile further, we combine the jet images with the kinematic features of our final state and finally observe a significant leap in the selection performance. In the end, we achieve a statistical significance of 4.2$\sigma$ for the fixed-$R$ case and 4.5$\sigma$ when the variable-$R$ jet clustering algorithm. It will then be interesting to note that our findings are  significantly better compared to the results obtained in \cite{Das:2018fog} for the $h_2\to ZZ\to$ `multi-lepton final' states, which provided less than 1$\sigma$ significance at the full Run 3 of the LHC. Thus, we will ultimately construct a VBF based access to $h_2\to h_1h_1\to b\bar bb\bar b$ chain decays, potentially competitive to the one based on ggF, as discussed in Refs.~\cite{Chakraborty:2023hrk,Hammad:2023sbd,Hammad:2025wst}. 

The organisation of this paper is as follows. We continue in Section~\ref{Sec:NMSSM} by introducing the theoretical framework of the NMSSM, paying particular attention to the structure of the Higgs sector and the production mechanisms that give rise to our signal process, i.e., $h_2\to h_1h_1\to b\bar bb\bar b$. Section \ref{Sec:collider}  details our approach to collider simulation and the conventional selection strategies that form the baseline for our VBF based analysis. Section \ref{Sec:results}  presents our main results, by comparing the performance of a cut-based analysis, to a image-based discrimination using a Multi-layer CNN architecture as well as the hybrid approach that combines the two. Finally, in Section \ref{Sec:summary}, we conclude.

%=======================================================
\section{The NMSSM} 
\label{Sec:NMSSM}

The NMSSM \cite{Fayet:1974pd,Dine:1981rt,Nilles:1982dy,Frere:1983ag,Derendinger:1983bz} extends the MSSM by incorporating an additional gauge singlet superfield (see \cite{Dedes:2000jp,Panagiotakopoulos:2000wp} for alternative realisations). As mentioned, this inclusion offers a dynamical solution to the 
discussed $\mu$-problem of the MSSM and accounts for the 125 GeV Higgs boson mass without requiring excessive fine-tuning of top squark masses. Beyond the MSSM Higgs doublet superfields $\hat{H}_u$ and $\hat{H}_d$, the NMSSM incorporates a gauge-singlet chiral superfield $\hat{S}$. This expansion further enriches the model particle spectrum, introducing extra Higgs and neutralino states that are absent in the minimal framework. The $Z_3$ invariant NMSSM superpotential can be written as  
\begin{equation} 
W_{\rm NMSSM} = W_{\rm MSSM} + \lambda \hat{H_u} \hat{H_d} \hat{S} + \frac{1}{3} \kappa \hat{S}^3, 
\end{equation}
where $W_{\rm MSSM}$ is the MSSM superpotential with the parameter $\mu$ set to zero. Once the scalar component of the  superfield $\hat{S}$ develops a Vacuum Expectation Value (VEV), $s\equiv\langle S\rangle$, it provides the effective term  $\mu_{\rm eff} = \lambda s$. Note that, in the absence of the term $\kappa$, the superpotential exhibits a Peccei-Quinn $U(1)_{PQ}$ symmetry \cite{Peccei:1977hh,Peccei:1977ur}. The dimensionless coupling $\kappa$ is specifically included to break this symmetry. However, the discrete $Z_3$ symmetry can be broken spontaneously without cosmological consequences (in terms of domain walls)~\cite{Zeldovich:1974uw,Panagiotakopoulos:1998yw,Dedes:2000jp}.

The soft SUSY breaking part of the Lagrangian is given by 
\begin{equation}
V_{\rm soft} = m^2_{H_u}|H_u|^2 + m^2_{H_d}|H_d|^2 + m^2_{S}|S|^2  + \lambda A_{\lambda} \hat{H_u} \hat{H_d} \hat{S} + \frac{1}{3} \kappa A_{\kappa}\hat{S}^3, 
\end{equation}
where $A_{\lambda}$ and $A_{\kappa}$ are free parameters that take positive and negative values. After EWSB in the CP-conserving case, the Higgs fields acquire non-zero VEVs, resulting in a physical Higgs sector comprising three CP-even bosons ($h_1, h_2, h_3$), two CP-odd bosons ($A_1, A_2$), and a charged pair ($H^\pm$). The additional terms in the NMSSM superpotential allow for greater control over the mass spectrum while maintaining perturbativity. At tree level, this sector is fully characterised by six parameters: $\lambda, \kappa, A_\lambda, A_\kappa$, $\mu_{\text{eff}}$ and $\tan\beta$ (the ratio of the two Higgs doublet VEVs). By adjusting the mixing between the doublet and singlet fields, any of the CP-even states can be identified as the 125 GeV SM-like Higgs boson. (Hereafter, we will take this state to be the $h_1$.)

In the NMSSM, the singlino component significantly modifies the neutralino mass matrix and the resulting phenomenology, especially the one of the DM candidate, the lightest neutralino (i.e., $\tilde{\chi}_1^0$). A singlino-like $\tilde{\chi}_1^0$ can satisfy stringent DM relic abundance and (in)direct detection bounds more easily than a pure MSSM-like neutralino. Furthermore, the reduced coupling of the singlino to the SM particles allows the NMSSM to also evade many existing collider bounds, thereby shifting the discovery potential toward a wider array of novel signatures in current and future runs of the LHC (including the HL-LHC). 

As already stated in the previous section, the focus of this work is to probe the Higgs sector of the NMSSM, specifically, the heavier Higgs boson $h_2$ produced through the VBF process and decaying into $h_1$ pairs, in turn yielding $b\bar bb\bar b$ as final decay products.  In most general cases, at the Leading Order (LO), the production of a Higgs boson (say, $h_2$) at the LHC is dominated by the ggF process, i.e.,  $g g \to h_2$. Other ways of producing the Higgs bosons are VBF and two associated production processes, namely, $q\to q\to Vh_2$, where $V = W/Z$, and $Q\bar{Q}H$, where $Q=b/t$. The dominance of one production process over the other primarily depends on the coupling of the relevant Higgs boson to the SM particles. For example, for lighter Higgses (say $m_{h_2} < 1$~TeV), the ggF mode dominates over all other processes with the associated production processes being the sub-dominant ones. However, as the mass of the Higgs boson increases, say $m_{h_2} \ge 1$~TeV, the VBF process starts to dominate over all others. 

%However, the dominance of one process over the other depends on the coupling of the Higgs boson with the SM particles. 
In the NMSSM, unlike the MSSM, the mixing between the Higgs superfields $\hat{H_u}$ and $\hat{H_d}$ with an additional singlet field $\hat{S}$ modifies the compositions of the physics Higgs states significantly and, therefore, affects their couplings with the SM particles as well. 
For example, one can find a parameter space region where, among the three CP-even neutral Higgs bosons, the lightest one (i.e., $h_1$) behaves like the observed SM-like Higgs boson, while the next-to-lightest one (i.e., $h_2$) could have a mass up to a TeV almost dominantly determined by the the VEVs of $\hat{S}$ and $\hat{H_d}$ fields. As a consequence, the couplings of the $h_2$ to the both the SM up- and down-type quarks will be highly suppressed and, furthermore, given the heavier squark masses and smaller $\tan\beta$ in the NMSSM, the contribution of the ggF process will be suppressed in general (as intimated). Hence, 
%Note that in the NMSSM, again, unlike the MSSM, by tuning the singlet component, we can suppress the $h_2 b\bar{b}$ coupling and, thus, the associated production with top/bottom quarks can be made negligible. In 
in this region of parameter space, the VBF can then be observed as the main production mechanism for probing the BSM Higgs boson $h_2$. In fact, the
decay profile of this heavier Higgs boson also shows some interesting features. As the coupling with the top/bottom quarks (and also the $\tau$ leptons) is suppressed, there exist only two modes through which the $h_2$ decays, namely, $h_2 \to VV$ (where $V = W/Z$), $h_2 \to h_1 h_1$ or $h_2 \to A_1 A_1$, provided that the corresponding mass differences allow for it. 
%This suppression occurs largely because the heavier Higgs state can assume a dominant singlet-like character.

Building upon the parameter space exploration conducted in Ref. \cite{Das:2018fog}, the present analysis focuses on a dedicated collider study centered around, the BP C therein, which implements the above faetures of the NMSSM parameter space. The input parameters for this BP, along with the mass spectrum for the SM-like and lighter BSM Higgs scalars, are summarised in Table \ref{tab:BP}. In Table \ref{tab:mixing}, we provide further details on reduced couplings of the $h_2$ scalar relevant for the primary production modes, specifically, ggF, VBF and associated production with $b$-quarks. Reduced couplings are denoted by $C_{h_2 X X}$ with $X$ being the SM particle under consideration with which the $h_2$ couples. In addition, we provide the relevant Branching Ratios (BRs) that display the decay signatures analyzed in the following section. Here we would like to note a few things: 1) the aforementioned BP has been obtained after a dedicated parameter space scan conducted using $\tt NMSSMTools$ \cite{Ellwanger:2004xm,Ellwanger:2006rn,Ellwanger:2005dv,Das:2011dg}; 2) the scan assumes that the squark masses are around 2.5 TeV while gauginos are  around 2 TeV; 
%along with satisfying the perturbative unitarity bounds on the input parameters, 
3) the constraints coming from the LHC Higgs data and flavour physics measurements are also satisfied using $\tt NMSSMTools$. For more details on the implementation of the parameter space scan and consideration of different experimental constraints, we refer to \cite{Das:2018fog}. 

\begin{table}[!htb]
    \centering
    \begin{tabular}{|c|c|c|c|c|c|c|c|}
    \hline 
        $\tan\beta$ & $\lambda$ & $\kappa$ & $A_\lambda$ (GeV)& $A_\kappa$ (GeV)& $\mu_{\rm eff}$ (GeV) & $m_{h_1}$ (GeV) & $m_{h_2}$ (GeV) \\ 
        \hline
         2 & 1.35 & 0.907 & 250 & $-520$ & 670 & 126.2 & 700 \\ 
    \hline
    \end{tabular}
    \caption{Input parameters for the BP considered in this analysis, along with the masses of the SM-like and the lightest CP-even BSM Higgs boson.}
    \label{tab:BP}
\end{table}

%%%%%%%%%%%%
\begin{table}[!htb]
    \centering
    \hspace*{-1.25cm}
    \begin{tabular}{|c|c|c|c|c|c|c|c|c|}
    \hline 
        $C^2_{h_2 t \bar{t}}$ & $C^2_{h_2 b \bar{b}}$ & $C^2_{h_2 \tau \tau}$ & $C^2_{h_2 VV}$ & $C^2_{h_2 gg}$ & ${\rm BR}({{h_2 \to WW}})$ & ${\rm BR}({{h_2 \to ZZ}})$ & ${\rm BR}({{h_2 \to t \bar{t}}})$ &${\rm BR}({{h_2 \to h_1 h_1}})$\\ 
        \hline
         0.005 & 0.59 & 0.59 & 0.044 & 0.005 & 0.363 & 0.17 & 0.01 & 0.45  \\ 
    \hline
    \end{tabular}
    \caption{Reduced couplings of the $h_2$ state to SM particles along with its largest BRs. }
    \label{tab:mixing}
\end{table}
%==================================================================

%=======================================================
\section{Event Simulation and Selection Strategy}  \label{Sec:collider}

In this section, we describe the simulation setup, event reconstruction procedure, and selection criteria employed to probe a heavy Higgs boson of mass $m_{h_2} = 700$ GeV produced via the VBF process at the 14 TeV LHC.

%-----------------------------
\begin{itemize}
\item {\underline{\bf Simulation details}}: 
We consider the production of a CP-even heavy Higgs boson with mass $m_{h_2} = 700$ GeV through the VBF process at the 14 TeV  LHC. The Higgs boson is produced in association with two forward/backward jets and subsequently decays into two SM-like Higgs bosons $h_1h_1$ with mass $m_{h_1} = 125$ GeV, each of which further decays into a pair of $b$-quarks, leading to a final state consisting of four $b$-jets and two light jets. The hard scattering process was generated using \texttt{MadGraph5\_aMC@NLO} \cite{Alwall:2014hca} (version 3.4.1) followed by \texttt{PYTHIA8} \cite{Sjostrand:2014zea} (version 8.310) for parton showering and hadronisation. The resulting HepMC output from \texttt{PYTHIA8} was subsequently passed to \texttt{Delphes} \cite{deFavereau:2013fsa} with the CMS detector card for fast detector simulation. We employ the \texttt{NNPDF23\_lo\_as\_0130\_qed} Parton Distribution Functions (PDFs) throughout the event generation, with both the factorization and renormalization scales set to $H_T/2$, where $H_T$ denotes the cumulative transverse energy of the colliding partons. 

For the signal events, we expect the presence of small radius forward/backward jets in association with the heavy Higgs boson $h_2$. Furthermore, given the large mass difference between the two Higgs bosons, namely $h_1$ and $h_2$, the lighter Higgs states are highly boosted and therefore are reconstructed as large-$R$(or fat) jets in the central region of the detector; these are the potential candidate for the SM-like Higgs boson states. Furthermore, the internal substructure of these boosted jets reveals the presence of smaller radius $b$-jets originating individually from the Higgs decay $h_1 \to b\bar{b}$. A precise reconstruction of both small and large-$R$ jets is essential for our analysis. While the small radius forward/backward jets play a crucial role in identifying the production topology and establishing the VBF-like structure of the event, the large-$R$ jets and their substructure are necessary to reconstruct the di-Higgs resonance and to efficiently discriminate the signal from relevant backgrounds. To this effect, we first cluster the {\tt EFLOW} objects using the \texttt{FastJet} library \cite{Cacciari:2011ma} with the anti-$k_T$ jet algorithm \cite{Cacciari:2008gp} employing a radius parameter of $R = 0.4$ and a minimum transverse momentum of $p_T = 20$ GeV. These form the set of small radius jets, which multiplicity is  denoted as $N_{j}$. These jets are then passed through a $b$-tagging algorithm. If the angular separation between the candidate $b$-jet and any of the $b$-hadrons is less than 0.4, then the jet is tagged as a $b$-jet. We further consider detector effects for flavour-tagging by imposing the $b$-tagging efficiency to be 70\% while using a $p_T$-dependent formula for the relevant mis-tagging probabilities \cite{ATLAS:2017bcq}. The multiplicity of these tagged $b$-jets are denoted by $N_{b}$. As the $b$-jets are the final products of the heavy Higgs boson decays, it is expected to be boosted in the central region of the detector. Therefore, events with at least 3 tagged 
$b$-jets with a minimum $p_T$ of  40 GeV and pseudorapidity $|\eta| \leq 2$ are selected for further analysis.  Those small-$R$ jets that are not tagged as $b$-jets having $|\eta| > 2$ and $p_T \geq 30$~GeV are considered as the  forward/backward jets of the event so long that such jets  have an invariant mass ($M_{jj}$) $\geq$ 700 GeV and a separation $|\Delta \eta_{jj}| \geq 4.5$ (according to the  VBF topology). 

In Figure \ref{fig:forwardjets}, we show the distribution of the angular separation of the two forward/backward jets $|\Delta \eta_{jj}|$ (left) and invariant mass of the same two jets $M_{jj}$ (right). The vertical line in each plot indicates the corresponding cut applied for the VBF selection. The plots also include the predictions for the dominant QCD background, due to $gg\to b\bar bb\bar b$ (hereafter, $4b$ events), in presence of Initial State Radiation (ISR) (which we will discuss in the next paragraph). 

%-------------------
\begin{figure}[!htb]
    \centering
    \includegraphics[width=0.45\linewidth]{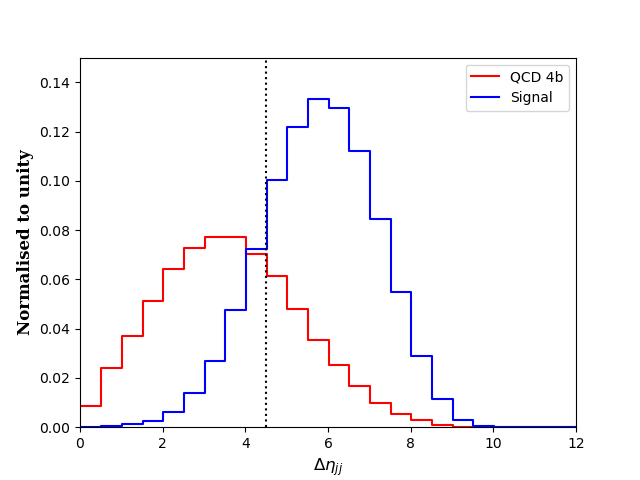}
    \includegraphics[width=0.45\linewidth]{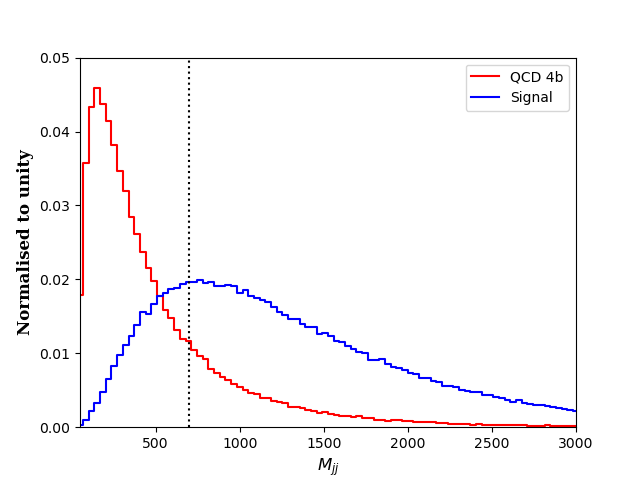}
    \caption{Normalised distributions of the pseudorapidity separation $\Delta \eta_{jj}$ (left panel) between the  two forward/backward jets and of their  invariant mass $M_{jj}$ (right panel). The vertical lines in each plot indicate the corresponding VBF selection cut applied, as listed in Table~\ref{tab:event-selection}.}
    \label{fig:forwardjets}
\end{figure}

%\textcolor{blue}{Now recalling the signal topology, we see, we had two jets in forward region and di-higgs(as the product of the heavy higgs) in the central region. Therefore, 
In addition to the above mentioned selection criteria, we also demand that each event must include at least two large-$R$ jets formed using the same {\tt EFLOW} objects with jet radius $R = 1.2$ and minimum transverse momentum $p_T > 200$ GeV in the central region (i.e., $|\eta| < 2)$. The multiplicity of large-$R$ jets is denoted as $N_J$. Note that these reconstruction parameters are optimised so as to capture both the overall event structure needed for the identification of the VBF topology and the substructure required for the reconstruction of the lightest two CP-even Higgs bosons. The details of the acceptance and selection cuts for the VBF topology are summarised in Table \ref{tab:event-selection}.

%------------------------------
\begin{table}[!b]
   \centering
    \begin{tabular}{c|c}
    \hline
     Acceptance cuts & VBF selection cuts\\ \hline
      $N_{j}\geq2,\;p_{T,j}\geq20~{\rm GeV},\;N_{b}\geq3,\;p_{T,b}\geq40~{\rm GeV}$,\;  & $|\eta_{j}|>2.0,\;|\Delta\eta_{jj}|\geq 4.5,\; M_{jj}\geq700$ GeV,\\  $|\eta_{b}|\leq2$,\;
     $N_{J}\geq 2,\; p_{T,J}\geq 200$ GeV&$p_{T,j}\geq30$ GeV \\ 
     \hline
    \end{tabular}
    \caption{The `acceptance' and `VBF selection' cuts applied to the generated events in our analysis.}
    \label{tab:event-selection}
\end{table}

%==========================================
\item {\underline{\bf Background analysis}}: 

The two most dominant SM backgrounds for the process under investigation are the QCD multijet events with four $b$-jets in the central region of the detector (the aforementioned $4b$ noise) and top quark pair production process ($t\bar{t}$)~\cite{ATLAS:2020jgy, Bishara:2016kjn}. We verified that the  $4b$ process constitutes approximately $95\%$ of the total background, while the contribution from $t\bar{t}$ events is around $5\%$. Furthermore, for the $t\bar{t}$ background, each top quark decays as $t \to bW$, followed by $W \to jj$, resulting in a characteristic three-prong jet substructure in the boosted regime. In contrast, the SM-like Higgs bosons in the signal events exhibit a two-prong substructure. Additionally, each top jet contains only one $b$-jet, making it significantly different from the signal topology. Therefore, requiring at least two $b$-jets within each of the two large-$R$ jets can effectively suppress $t\bar{t}$ background. Hence, in the remainder (as intimated), we neglect $t\bar t$ events from the SM background dataset. 

In the case of the QCD multijet events, the signal-like topology requires four $b$-jets in the central region along with at least two forward/backward jets (from ISR) to mimic the VBF topology. However, generating such high jet multiplicity events is computationally expensive. To address this problem, we follow the strategy of Ref.~\cite{Bishara:2016kjn}, where the hard process is generated with four $b$-quarks at the matrix element level and subsequently passed through {\tt PYTHIA} to obtain additional forward/backward jets, thereby approximating the VBF topology. The process $pp \to 4b$ receives contributions from both the $q\bar q \to 4b$ and $gg \to 4b$ channels. However, the $gg \to 4b$ channel dominates the cross section, contributing nearly $99\%$ of the total. Therefore, we focus exclusively on $gg \to 4b$ for simulating the $4b $ background events. 

To further ensure that  background events closely resembles the signal topology, we impose a set of generation-level cuts in the {\tt MadGraph} run card, as outlined below:
\begin{itemize}
    \item $\sqrt{\hat{s}} = 700~\text{GeV}, \quad H_T \geq 600~\text{GeV}$,
    \item $|\eta_b| \leq 3, \quad p_{T,b} \geq 30~\text{GeV}$,
    \item $M_{bb} > 90~\text{GeV}$,
\end{itemize}
where  $\sqrt{\hat{s}}$ is the centre-of-mass energy at the parton level, $H_T$ is the scalar $p_T$ sum of all the visible objects present in the hard process, $|\eta_b|$ and $p_{T,b}$ are the pseudorapidity and transverse momentum of the $b$-quarks and finally $M_{bb}$ is the invariant mass of the any  $b$-quark pairs present in the final state. Note that these cuts are optimised to enhance the likelihood that the two large-$R$ jets in the central region of the background process reconstruct an invariant mass close to that of the heavy Higgs boson $h_2$. With this selection, the invariant mass distribution peaks around $700~\text{GeV}$ and the resulting cross section after all cuts is approximately $1240~\text{fb}$
at the Leading Order (LO) in $\alpha_s$, the strong coupling constant. The production cross section (at the same perturbative level) for the signal events with $m_{h_2} = 700$ GeV is 10 fb. However, the actual cross section for the signal process after multiplying by all the relevant BRs, namely, ${\rm BR}(h_2 \to h_1 h_1)$ = 45\% and ${\rm BR}(h_1 \to b \bar{b})$ = 60\% (for both the light Higgses), becomes 1.62 fb.

%------------------

\end{itemize}

%=============================================================

%=====================================================================
\section{Results} \label{Sec:results} 

To evaluate the performance of our jet identification framework, we adopt a multiple strategy that progressively incorporates higher levels of complexity. We begin with a baseline cut-and-count method utilizing standard reconstructed observables to establish a performance floor. To improve upon this, we employ a DL approach using a Multi-layer CNN architecture designed to capture the energy distribution within the large-$R$ jets. The final stage of our refinement involved augmenting this CNN framework with high level kinematic features, thereby creating a hybrid model that leverages information stored in both low  and high level inputs. Throughout this study, the robustness of each method is tested using two different jet clustering algorithms: a standard fixed-$R$ algorithm (the aforementioned anti-$k_T$ one) and the variable-$R$ algorithm with radius parameter varying from 0.4 to 2.0 (following Ref.~\cite{Chakraborty:2023hrk}), in turn allowing us to assess the influence of jet boundaries on the classification accuracy.

The events that are passed through the above-mentioned selection cuts (as outlined in Table \ref{tab:event-selection}) must also satisfy the following requirement: each of the large-$R$ jets must  include two small-$R$ $b$-tagged jets of radius 0.4 reconstructed using the constituents of the large-$R$ jet. The same b-tagging procedure is adopted as discussed in the previous section. This additional criterion improves the purity of the signal events and also helps to reduce background contamination. The efficiencies, calculated as the percentage of events that passed the cuts over the total simulated events, of these cuts on the signal and the SM background are listed in Table \ref{tab:efficiency}. 

%-------------------------
\begin{table}[htb!]
    \centering
    \hspace{-2cm}
    \begin{tabular}{|c|c|c|}
    \hline
    Selection cuts    &   & Efficiency (in \%)\\ \hline
     Acceptance cuts + VBF selection &  Signal       & 10.17  \\ \cline{2-3}
     %& S3: Acceptance cuts(+$N_{jet}\geq2$+$N_{j}^{b}\geq$2)+$M_{bb}=(100,150), M_{jet-jet}=(600,800)$+VBF& 4 \\ \hline
     
%    +\sout{$N_{J}\geq2$}+$N_{J}^{b}\geq$2 &  Background & 0.45 \\ 
     + $N_{J}^{b}\geq$2 &  Background & 0.45 \\ 
     %     & S3: Acceptance cuts(+$N_{jet}\geq2$+$N_{j}^{b}\geq$2)+$M_{bb}=(100,150), M_{jet-jet}=(600,800)$+VBF& 0.03 \\ 
     \hline
    \end{tabular}
\caption{The selection of cuts applied to the signal and background events with their respective efficiencies. The quantity $N_{J}^{b}$ denotes the number of b-tagged subjet for a given large-$R$ jet (J).  }
    \label{tab:efficiency}
\end{table}
%--------------------------------------

%$\bullet$ {\underline{\bf Simple Cut-and-Count method}}: \\  
\subsection{Simple Cut-and-Count Method}
One of the simplest methods of separating the signal from the background events would be to impose an additional cut around the mass of the CP-even light and heavier Higgs bosons, namely, to demand, e.g.,  $M_{bb} \sim M_{h_1} = [100 - 150]$~GeV and $M_{4b} \sim M_{h_2} = [600 - 800]$~GeV. These cuts significantly reduce the contribution of the SM background events and achieve statistical significance around 1.7$\sigma$ for 300 fb$^{-1}$ of integrated luminosity, as shown in Table \ref{tab:signi-cuts}. Hereafter, we use the following formula for calculating the signal significance ${\cal S}$ \cite{Cowan:2010js}: 
\begin{equation}
\mathcal{S}  = \sqrt{2 \left[ (S + B) \ln\left(1 + \frac{S}{B}\right) - S \right]},
\end{equation}
with S and B being the number of signal and background events, respectively,  that satisfy the various selection cuts imposed. 

%--------------------------
\begin{table}[htb!]
    \centering
    \begin{tabular}{|c|c|c|c|}
    \hline
       Process& Cross section$\times$BR (fb)  & Efficiency & Number of events \\\hline
       Signal(S)  & 1.62 & 0.04 & 19.4\\\hline
       $pp\to bbbb$ (B)  & 1240 & 0.0003 &111.6\\ \hline %\hline 
%       S/B &&& 0.174\\\hline
%       $\mathcal{S}$ &&& 1.7  \\\hline
    \end{tabular}
    \vskip 0.3cm 
        \begin{tabular}{|c|c|c|c|}
        \hline
      S & B &  S/B & $\mathcal{S}$ \\\hline
      19.4 & 111.6 & 0.174 & 1.79 \\\hline
    \end{tabular}
    \caption{Cut efficiency, Signal (S) and Background (B) rates as well as significance (${\cal S}$) after applying the basic acceptance and selection cuts along with an additional selection cut around the heavy Higgs boson mass  ([600 GeV,800 GeV]) as well as around the SM Higgs boson mass ([100 GeV,150 GeV]),  using the integrated luminosity $\mathcal{L}=300$ fb$^{-1}$.}
    \label{tab:signi-cuts}
\end{table}

%--------------------------------------------------------------
%---- jet images + CNN 
%$\bullet$ {\underline{\bf Multi-Scale CNN architecture for Jet Images only}}: \\ 
\subsection{Multi-Layer CNN Architecture for Jet Images Only} 
\begin{figure}[htb!]
    \centering
    \includegraphics[width=0.9\linewidth]{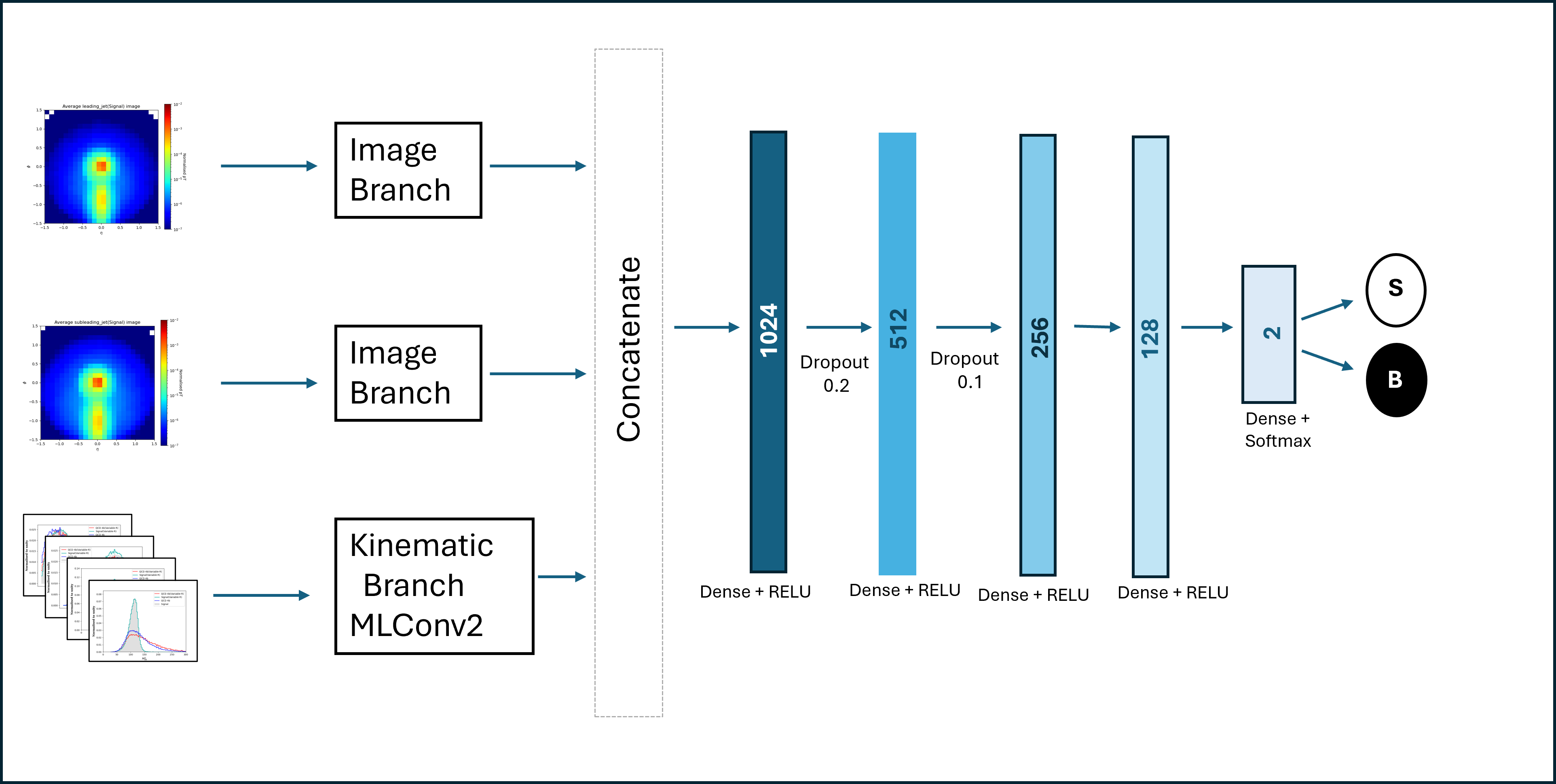}
    \caption{Schematic illustration of the DL architecture employed in this work for classifying signal  against background events.}
    \label{fig:dnn}
\end{figure}

\begin{figure}[htb!]
    \centering
    \includegraphics[width=0.45\linewidth]{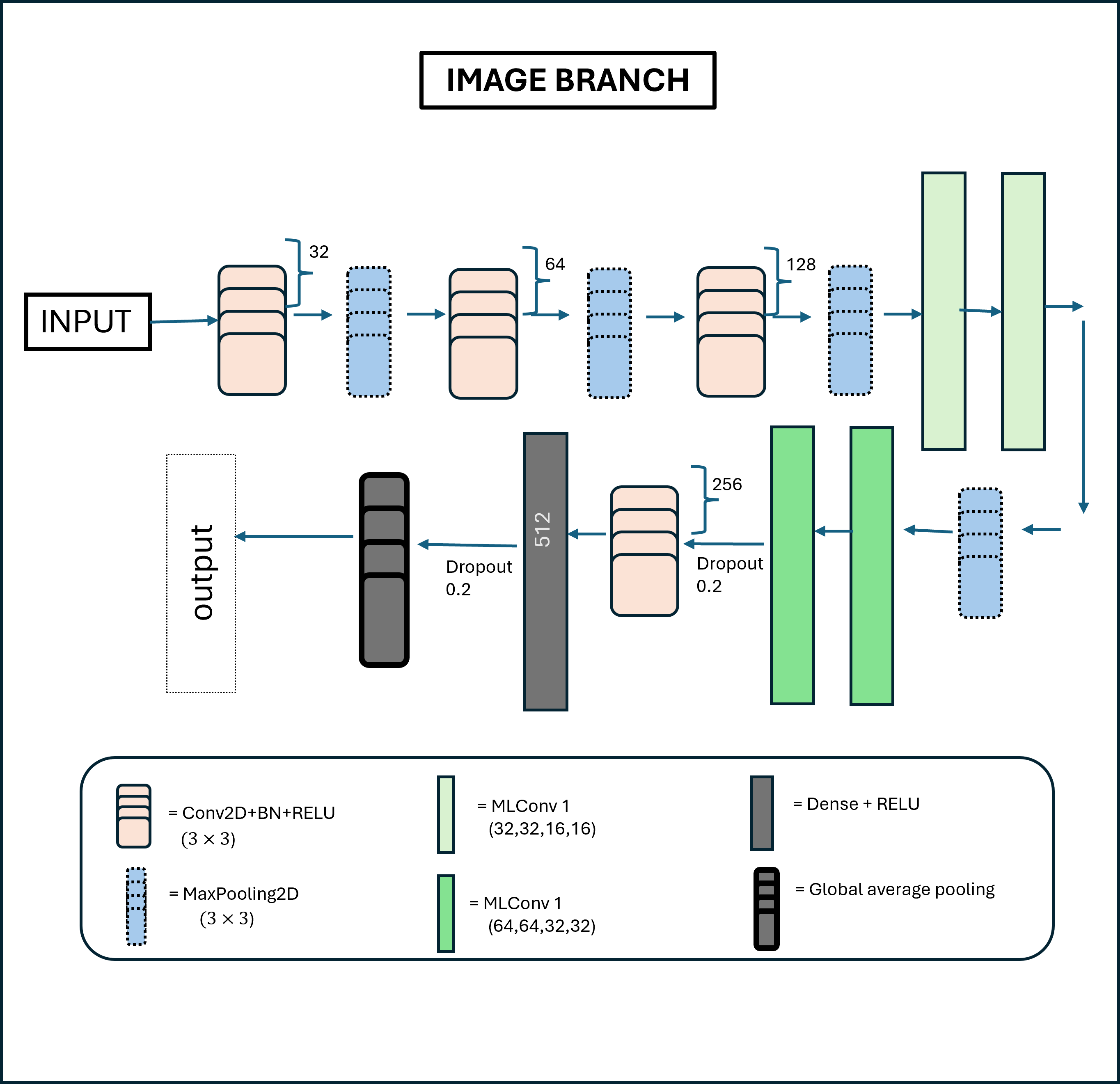}
    \includegraphics[width=0.45\linewidth]{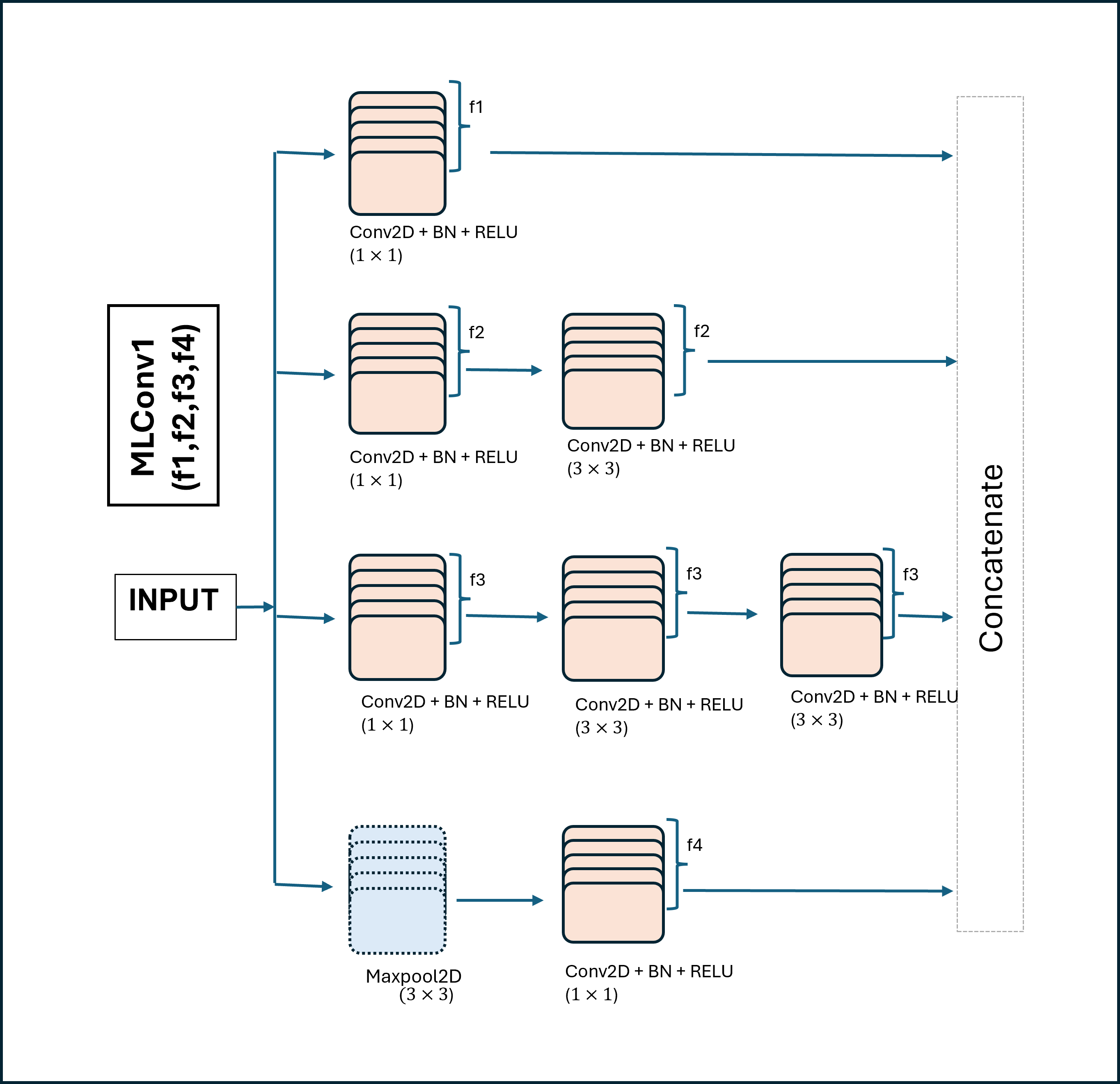}
    \caption{The explicit representation of the image branch mentioned in Figure \ref{fig:dnn} (left panel). The Multi-layer CNN block used in the image branch is also illustrated (right panel).}
    \label{fig:MSconv1}
\end{figure}

For the signal, each event contains two boosted SM-like Higgs boson with mass around 125 GeV, and therefore, we expect there will be two large-R jets each representing the SM Higgs boson. Therefore, we prepare jet images for the leading and sub-leading large-$R$ jets, which are then passed in parallel through an image branch composed of four Multi-layer CNN architecture, interleaved with convolutional blocks, ultimately yielding a high-dimensional latent feature vector from an intermediate dense layer (see Figure~\ref{fig:MSconv1}). While designing the Multi-layer CNN architecture, we take inspiration from the {\tt Google Inception network} ({\tt GoogLeNet}) \cite{szegedy2015going, szegedy2016rethinking}, which is widely used to extract features at multiple depths of a generic CNN. The resulting latent feature vectors are then concatenated and fed into a series of dense layers, as illustrated in Figure \ref{fig:dnn} (excluding the kinematic branch, which we will discuss in the following), to classify events as signal- or background-like. 

%------------------------------

\begin{figure}[htb!]
    \centering
    \includegraphics[width=0.45\linewidth]{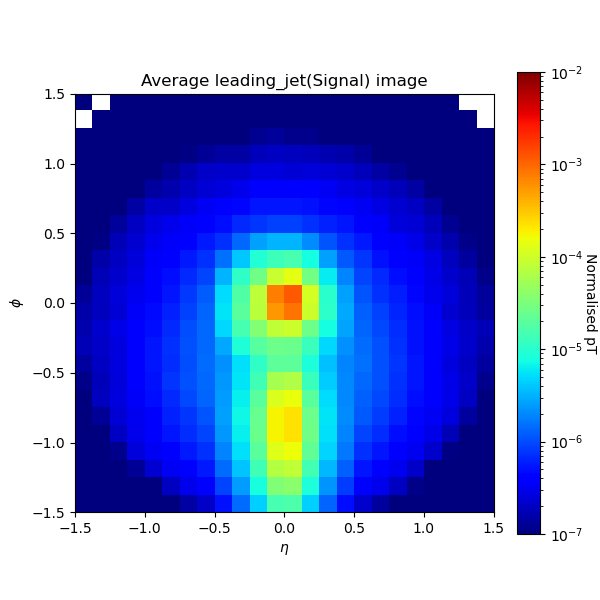}
    \includegraphics[width=0.45\linewidth]{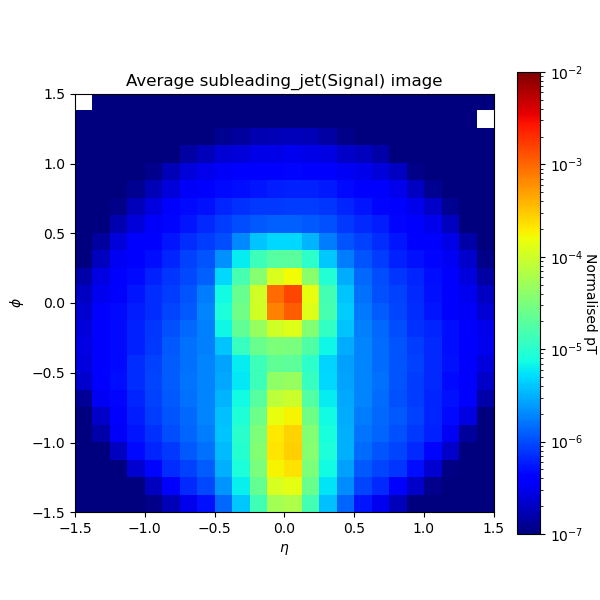}
     \includegraphics[width=0.45\linewidth]{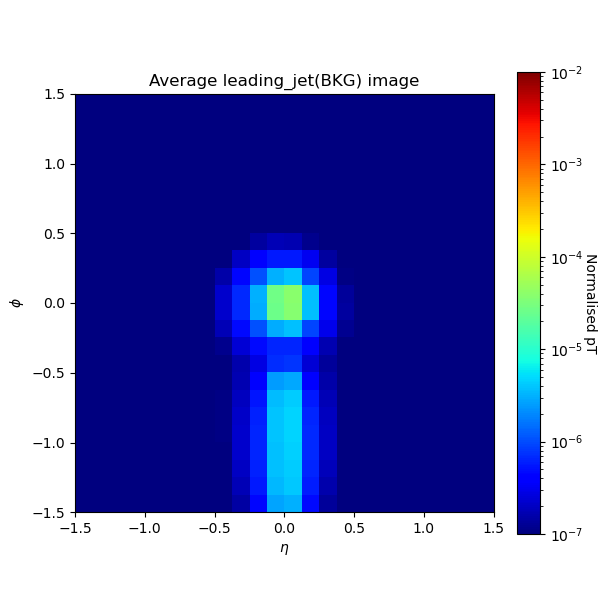}
    \includegraphics[width=0.45\linewidth]{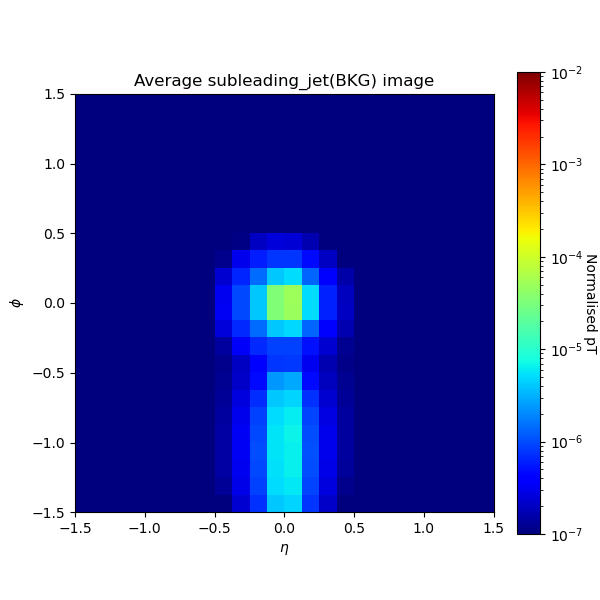}
    \caption{The average jet image of signal events for the leading (top left) and sub-leading (top right) jets for the fixed-$R$ case. In the bottom row, we display the same but for background events. }
    \label{fig:jetimageFixedR}
\end{figure}

In Figure \ref{fig:jetimageFixedR}, we display the average images of the leading and sub-leading jets for the signal (top row) and background (bottom row) events considering a total of  300,000 events. These images were formed using the constituents of the large-$R$ candidate jets\footnote{We follow \cite{Barnard:2016qma} for the full jet image construction and preprocessing steps.}. Note that, in this analysis, we do not impose the two mass window cuts for these candidate jets, rather we allow the Multi-layer CNN algorithm to learn the internal kinematic profile to build the classifier.  

%---------------------

%--var R 

The fixed jet radius parameter $R$ determines the angular boundary of a jet in the pseudorapidity-azimuthal angle plane. However, its effectiveness sometimes become limited by the fact that high-$p_T$ particles produce narrow cones while lower-$p_T$ objects spread over wider angles. To address this, the variable-$R$ jet clustering algorithm \cite{Krohn:2009zg} emerged, by adopting a variable jet radius in each event, thereby challenging the traditional fixed-$R$ schemes. By dynamically scaling the radius based on the momentum of the constituents, this approach ensures that the underlying physics is captured more precisely than by a fixed-$R$ method. To maximise the precision of the reconstructed resonance mass peaks, we optimise the choice of the parameter $\rho$. For our variable-$R$ algorithm, we set $\rho = 400$ with $R_{\rm min} = 0.4$ and $R_{max} = 2.0$, a configuration informed by both the $p_T$ scale of fixed-cone $b$-jets and a systematic scan of the $\rho$ values\footnote{This choice is thus different from the case of Ref.~\cite{Chakraborty:2023hrk}.}. 

\begin{figure}[htb!]
    \centering
    \includegraphics[width=0.45\linewidth]{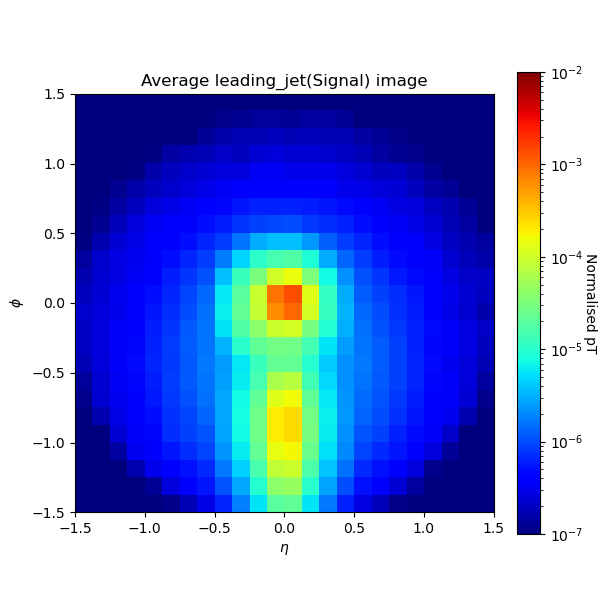}
    \includegraphics[width=0.45\linewidth]{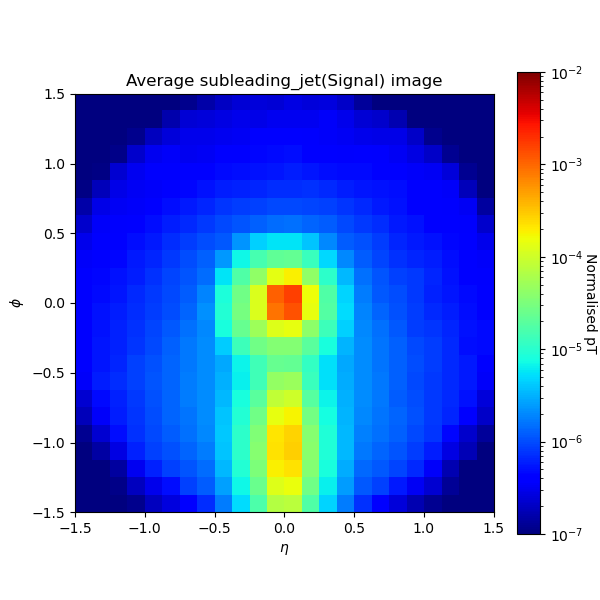}
     \includegraphics[width=0.45\linewidth]{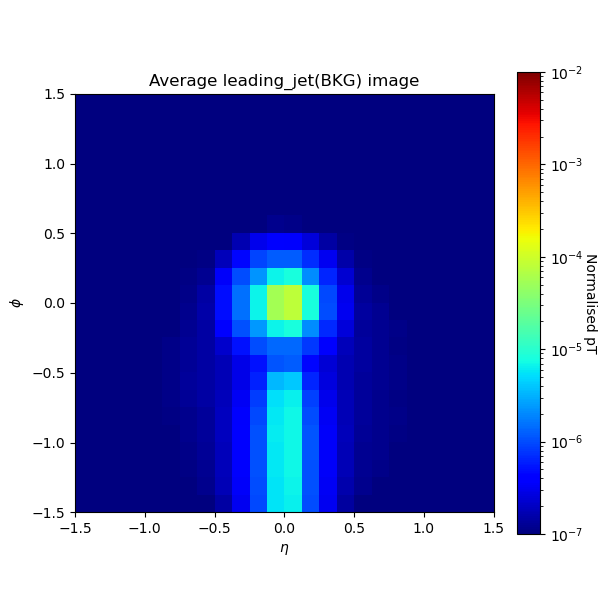}
    \includegraphics[width=0.45\linewidth]{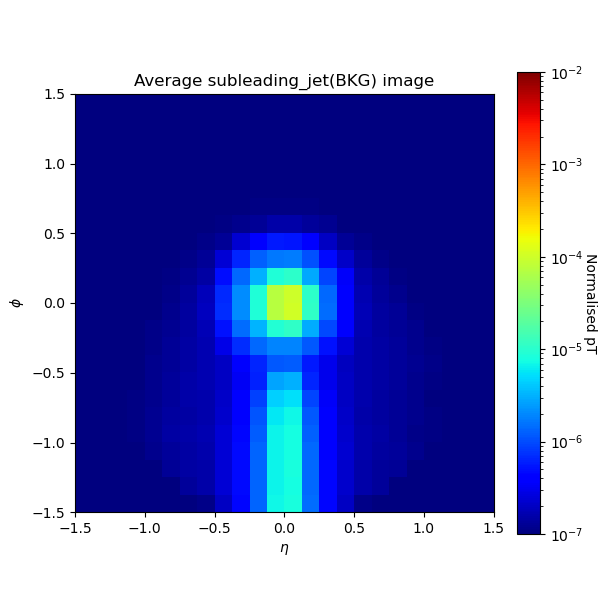}
 \caption{The average jet-image of  signal events for the leading (top left) and subleading (top right) jets for the variable-$R$ case (with $\rho=400$). In the bottom row, we display the same but for background events.}
    \label{fig:jetimageVarR}
\end{figure}

Following the fixed-$R$ analysis, we prepare the average images of the leading and sub-leading jets using the constituents of these jets obtained applying the variable-$R$ jet algorithm with 450,000 events for signal and background, as shown in Figure \ref{fig:jetimageVarR}. 
%----------------------

For each choice of jet clustering algorithm,  the two jet-images (leading and sub-leading) are then analysed through the Multi-layer CNN algorithm individually and then combined to classify the events as signal or background. As Table \ref{tab:2CNN-fixedR} suggests, even without the Higgs mass window cuts, the approach achieves better significance ($\sim 2.8 \sigma$) than the cut-and-count based one. In fact, the results are better 
(${\cal S} \sim 2 \sigma$) than the traditional approach when a systematic uncertainty of 10\% \ is considered. The same exercise was performed for the variable-$R$ case, however, no significant differences were observed in the performance of the Multi-layer CNN-based trained model, as tabulated in Table \ref{tab:2CNN-varR}.

%------------------------------
\begin{table}[htb!]
    \centering
   \begin{tabular}{|c|c|c|c|}
    \hline
       Process& Cross section$\times$BR (fb)  & Efficiency & Number of events \\\hline
       Signal (S)  & 1.62 & 0.0479 & 23.3\\\hline
       $pp\to bbbb$ (B)  & 1240 & 0.00016 & 59.7\\ \hline %\hline 
%       S/B &&& 0.174\\\hline
%       $\mathcal{S}$ &&& 1.7  \\\hline
    \end{tabular}
    \vskip 0.3cm 
        \begin{tabular}{|c|c|c|c|}
        \hline
      S & B &  S/B & $\mathcal{S}$ \\\hline
      23.3 & 59.7 & 0.39 & 2.8 \\\hline
    \end{tabular}   
    \caption{Cut efficiency and signal significance after applying the basic acceptance cuts and optimisation of the signal efficiencies using the (double) Multi-layer CNN architecture at the integrated luminosity $\mathcal{L}=300$ fb$^{-1}$ for the case of the fixed-$R$ jet clustering algorithm.}
    \label{tab:2CNN-fixedR}
\end{table}

%-------------------------

\begin{table}[htb!]
    \centering
   \begin{tabular}{|c|c|c|c|}
    \hline
       Process& Cross section$\times$BR (fb)  & Efficiency & Number of events \\\hline
       Signal (S)  & 1.62 & 0.0449 & 21.8\\\hline
       $pp\to bbbb$ (B)  & 1240 & 0.00015 & 55.8\\ \hline %\hline 
%       S/B &&& 0.174\\\hline
%       $\mathcal{S}$ &&& 1.7  \\\hline
    \end{tabular}
    \vskip 0.3cm 
        \begin{tabular}{|c|c|c|c|}
        \hline
      S & B &  S/B & $\mathcal{S}$ \\\hline
      21.8 & 55.8 & 0.39 & 2.8 \\\hline
    \end{tabular}   
    \caption{Cut efficiency and signal significance after applying the basic acceptance cuts and optimisation of the signal efficiencies using the (double) Multi-layer CNN architecture at the integrated luminosity $\mathcal{L}=300$ fb$^{-1}$ for the case of the variable-$R$ jet clustering algorithm.}
    \label{tab:2CNN-varR}
\end{table}

%\newpage
\vskip 0.5cm 
%------------------------------------------------------
%$\bullet$ {\underline{\bf Multi-Scale CNN architecture for Jet Images and Kinematic features}}: \\ 
\subsection{Multi-Layer CNN Architecture for Jet Images with Kinematic Features} 

\begin{figure}[htb!]
    \centering
     \includegraphics[width=0.5\linewidth]{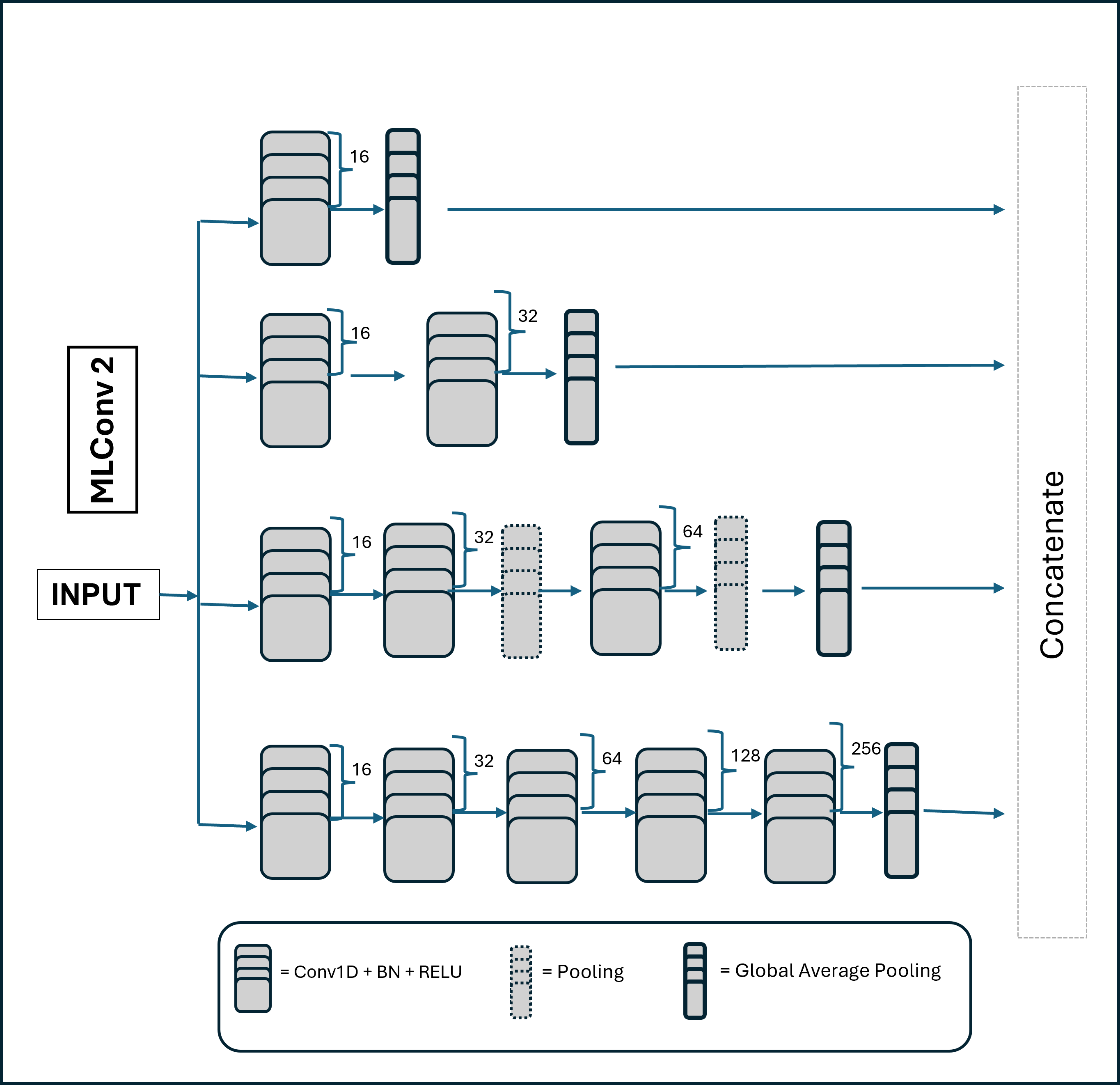}
    \caption{The kinematic branch of Figure \ref{fig:dnn} that uses one Multi-layer CNN block. }
    \label{fig:MSconv2}
\end{figure}

Realizing the importance of high-level objects in the classification of objects along with jet constituent-based inputs, as highlighted in many previous works, e.g., \cite{Hammad:2023sbd,Hammad:2025wst}, we extend the aforementioned architecture by incorporating event level kinematic information. The kinematic inputs, as listed in Table \ref{tab:variables}, are processed through a separate Multi-layer CNN branch, as shown in Figure \ref{fig:MSconv2}. The output of this branch is concatenated with the latent feature vectors obtained from the image branches. The combined representation is subsequently passed through the same dense layers as before for final classification. 

%----------------------------

%%%% kinematic variables ... 
We first apply the basic acceptance and selection cuts to both signal and  background, as adopted in the previous two methods, for both jet clustering algorithms. Before we discuss our findings, let us briefly outline the list of variables (see Table \ref{tab:variables}) constructed using the kinematic information of the event below.
 
\begin{figure}[!htb]
    \centering
    \includegraphics[width=0.32\linewidth]{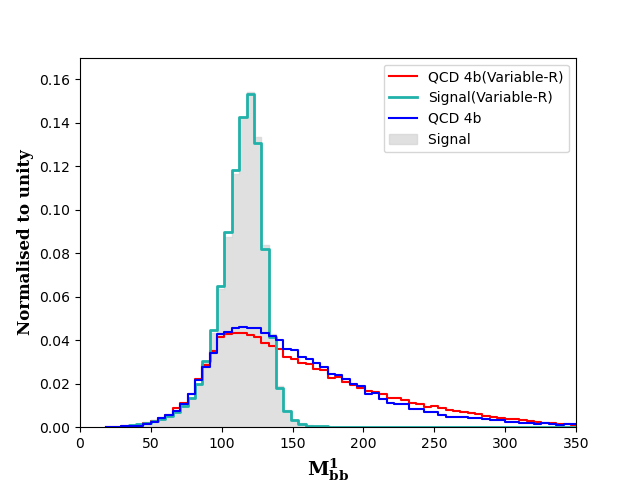}
    \includegraphics[width=0.32\linewidth]{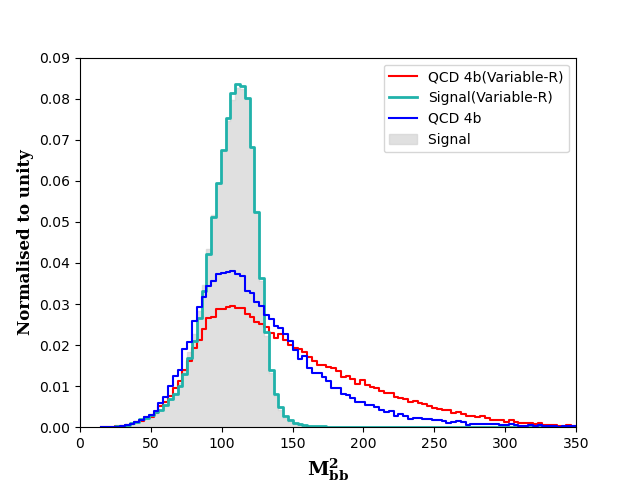}
    \includegraphics[width=0.32\linewidth]{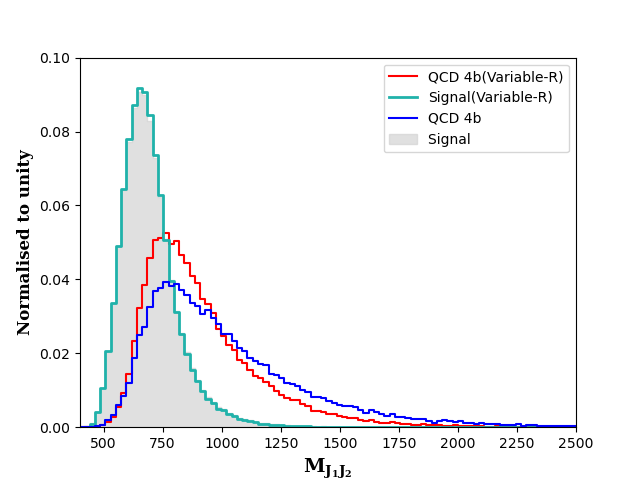}
    \caption{Normalised distributions for the invariant mass of the two $b$-tagged jets corresponding to the leading and sub-leading large-$R$ (Higgs candidate) jets. The right most plot shows the invariant mass of the the two leading large-$R$ jets. All the plots are drawn for the signal and background events with two choices of jet clustering algorithms. }
    \label{fig:bb}
\end{figure}

\begin{figure}[!htb]
    \centering
    \includegraphics[width=0.35\linewidth]{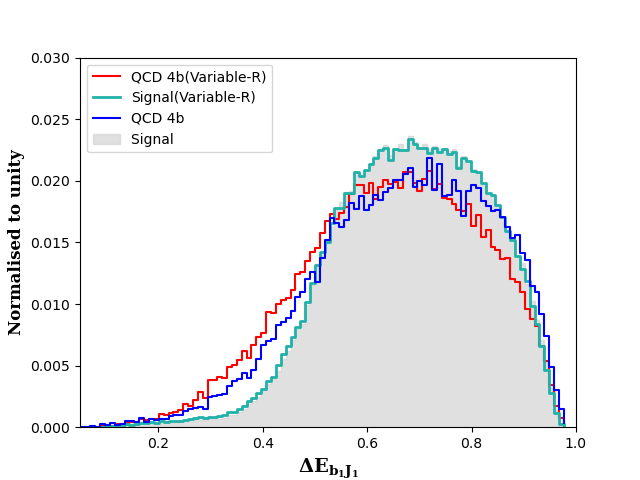}
    \includegraphics[width=0.35\linewidth]{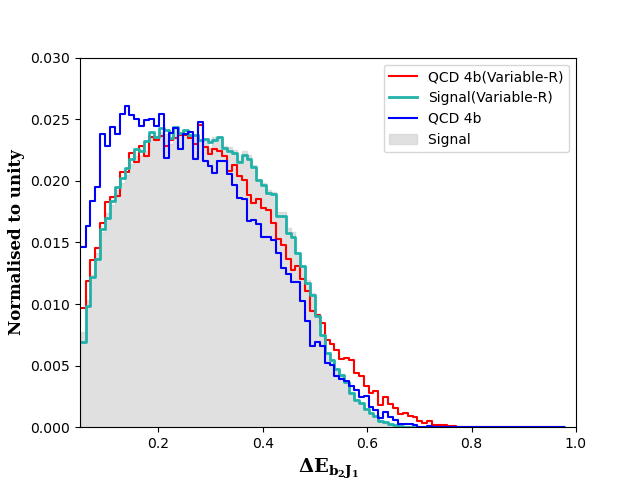}
 \caption{Normalised distributions for the energy fraction of the
  leading (left panel) and sub-leading (right panel) $b$-tagged subjets with respect to the leading large-$R$ jet. }
    \label{fig:Ebb}
\end{figure}

\begin{figure}[!htb]
    \centering
    \includegraphics[width=0.32\linewidth]{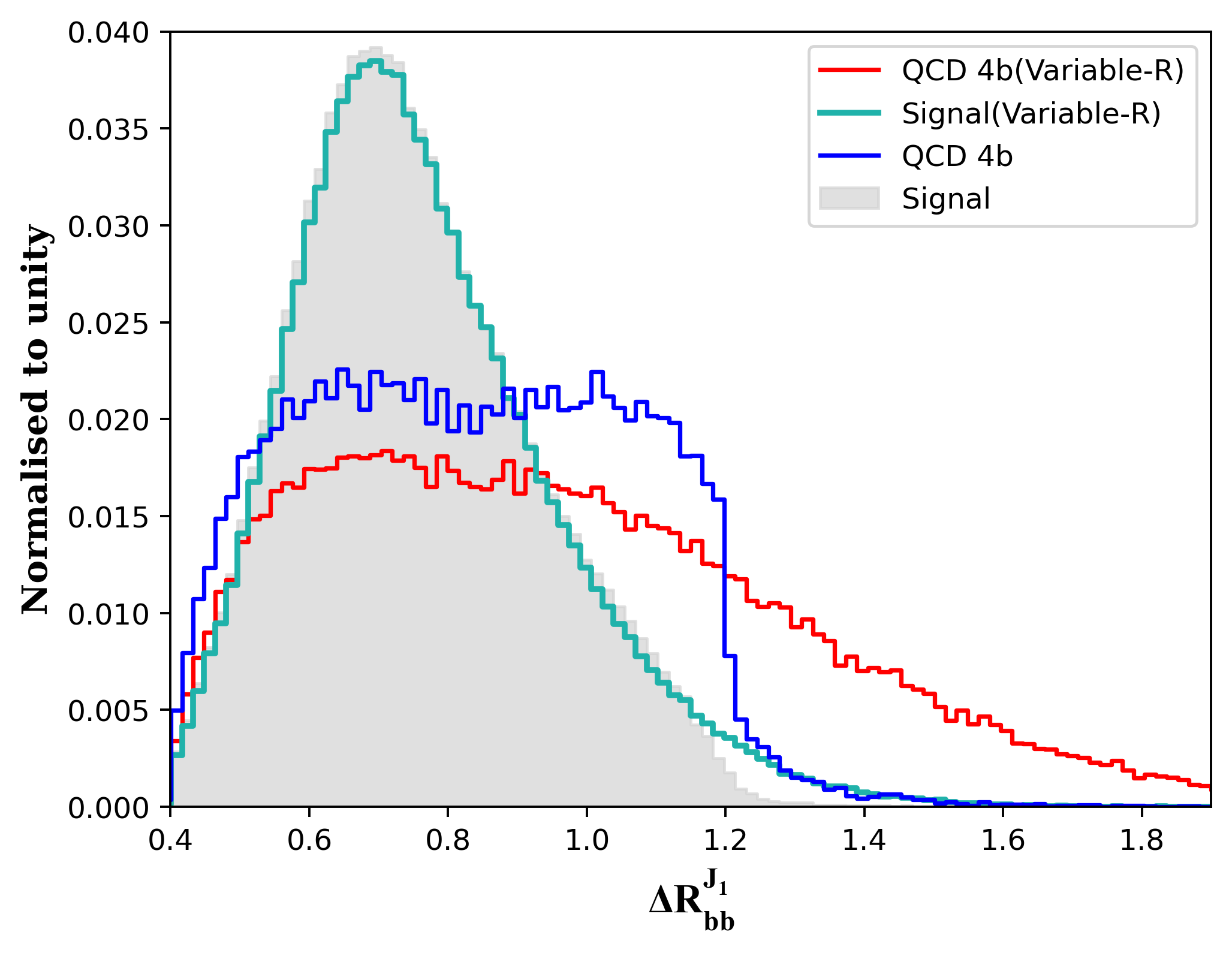}
    \includegraphics[width=0.32\linewidth]{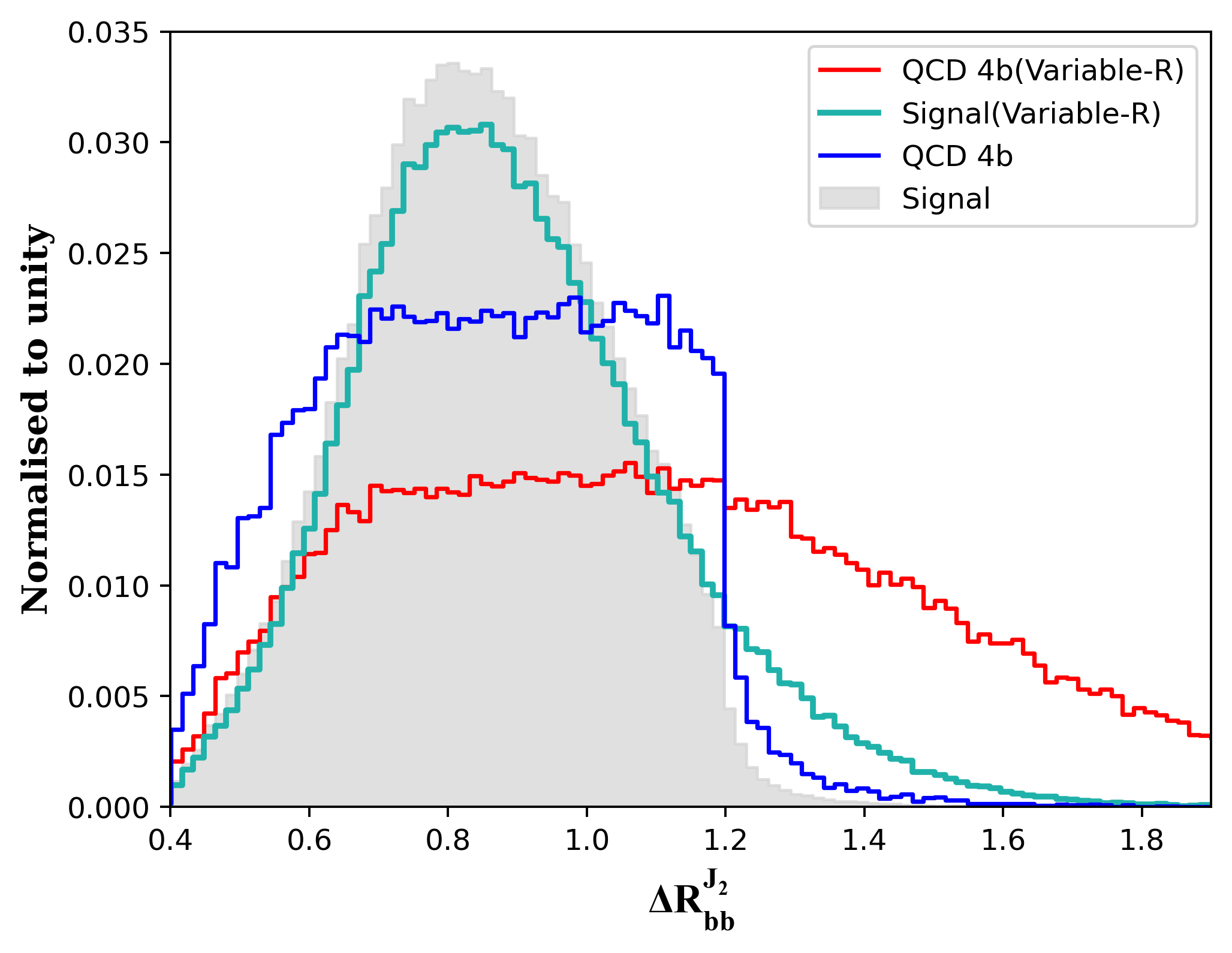}
    \includegraphics[width=0.32\linewidth]{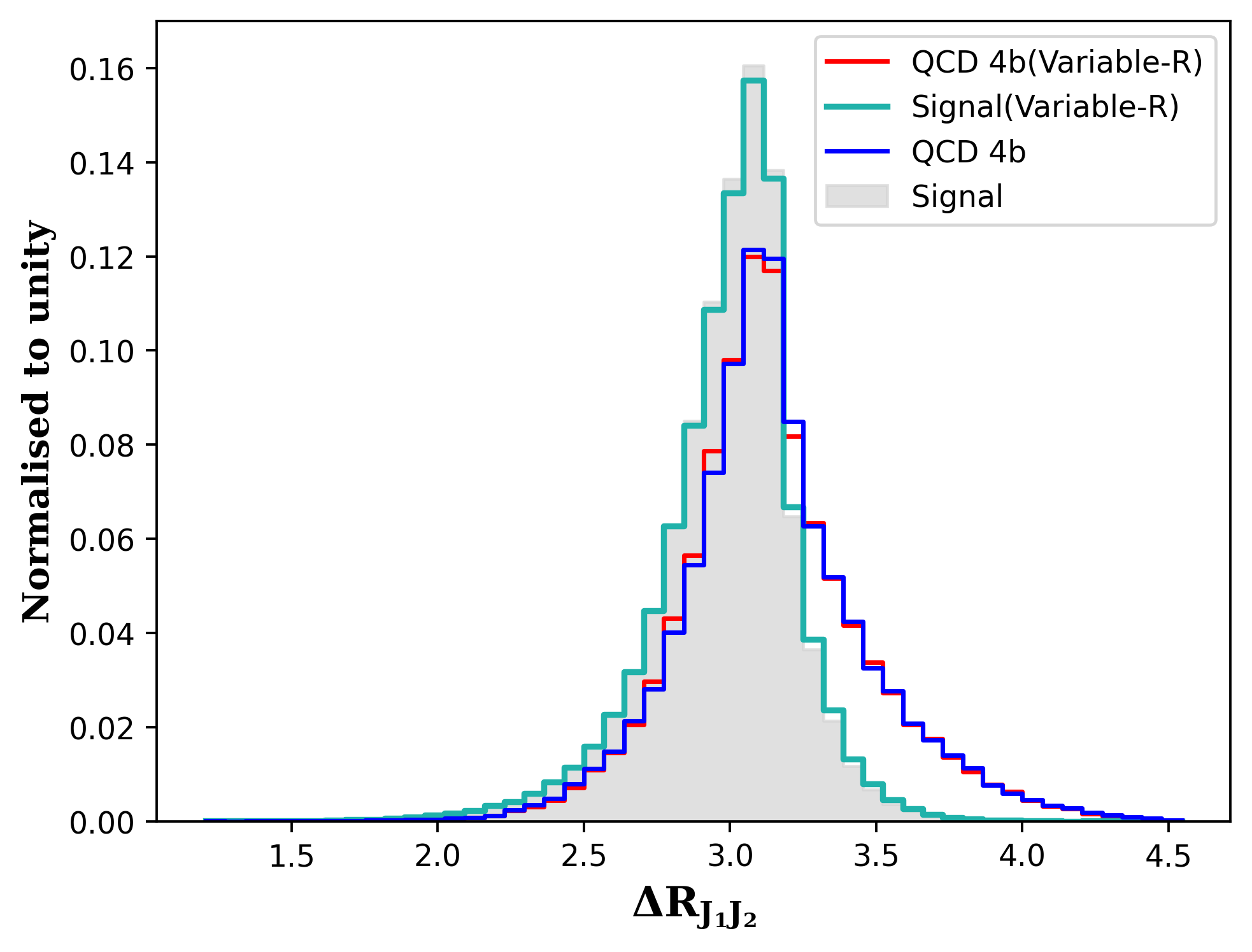}
 \caption{Normalised distributions for the angular separation ($\Delta R$) in the ($\eta, \phi$) plane between the two $b$-tagged subjets of the leading and sub-leading large-$R$ jets, shown in the left and central panel, respectively. In the right panel, we display the distribution of the angular separation between the two leading large-$R$ jets.}
    \label{fig:delRbb}
\end{figure}

\begin{itemize}
    \item $M_{J_1J_2}$: Invariant mass of the leading and sub-leading large-$R$ jets. The distribution, shown in the right-most plot of Figure \ref{fig:bb}, peaks at the mass of the heavy resonance for the signal events, as expected.  

    \item $M^{i}_{bb}$ ($i = 1,2$): Invariant mass of the two $b$-tagged sub-jets for each of the two leading large-$R$ jets: see the first two plots of Figure \ref{fig:bb}. Both distributions confirm the presence of a Higgs boson of mass around 125 GeV inside each of the large-$R$jets (for signal events). 
    
    \item $\Delta E_{b_i J_j}$ ($i,j = 1,2$): The energy fraction of the ${\it j^{th}}$ large-$R$ candidate jet carried by the ${\it i^{th}}$ sub-jet (inside the former) that is tagged as a $b$-jet. In Figure \ref{fig:Ebb}, we show the distributions of this variable, where the left panel for the energy fraction carried by the leading and the right panel for sub-leading $b$-jet of the leading large-$R$ jet, denoted as $\Delta E_{b_1J_1}$ and $\Delta E_{b_2J_1}$, respectively. We calculate the same for the other sub-leading large-$R$ jet ($J_2$) and use those as input for the kinematic layer.
    
    \item $\Delta R^{J_i}_{bb}$ ($i = 1,2$) and $\Delta R_{J_1J_2}$: The angular separation between the $b$-tagged subjets of the two Higgs candidate large-$R$ jets, denoted by $\Delta R^{J_1}_{bb}$ and $\Delta R^{J_2}_{bb}$, are shown in the first two plots of Figure \ref{fig:delRbb}. The distributions clearly illustrate that, for signal events, they peak at 0.7 and 0.8, respectively, depicting the two $b$-jets inside the Higgs candidate jet as being well separated, whereas, for the background, these variable show an almost flat distribution, as, unlike the signal events, the background ones are not coming from resonant objects. The angular separation between the Higgs candidate jets, denoted by $\Delta R_{J_1J_2}$ and shown in the right-most plot of Figure \ref{fig:delRbb}, is well separated for the signal events, as the heavy resonance, i.e., $h_2$ (with a mass of 700 GeV) is produced almost at rest. A longer tail for the QCD events again confirms the absence of a resonance in the simulated process, however, the choice of generation level cuts allows us to cover the phase space having a significant overlap with the signal regions.  
\end{itemize}

%---------------
As a quick summary, we tabulate all the variables used in the kinematic layer in Table \ref{tab:variables}. The last two variables, namely $y_{J_1}$ and $y_{J_2}$, are the measure of the symmetry among the two $b$-tagged subjets $j_1$ and $j_2$ of a candidate Higgs jet $J$, defined as $y_J = \frac{\min(p_{T,j1}^2, p_{T,j2}^2)}{m_J^2} \Delta R_{j1,j2}^2$. For signal events, the subjets are expected to be relatively symmetric, unlike the background events, due to the absence of any resonance in the latter. Furthermore, the distributions of all the kinematic variables are also drawn for the case of the variable-$R$ jet clustering algorithm with $\rho = 400$ and are overlayed to the fixed-$R$ case. Some minor but noticeable differences are observed for some of the kinematic features, which play an important role in improving the results when we consider these kinematic observables as inputs (as we shall see below). 

%---------------------------------
\begin{table}[!tbh]
    \centering
    \begin{tabular}{|c|}
    \hline
        Kinematic variables \\\hline
        \\
         $M_{J_1J_2},\;M^1_{bb},\;M^2_{bb},\;\Delta E_{b_1J_1},\;\Delta E_{b_2J_1},\;\Delta E_{b_1J_2},\;\Delta E_{b_2J_2},\;\Delta R^{J_1}_{bb},\;\Delta R^{J_2}_{bb},\;\Delta R_{J_1J_2},\;y_{J_1},\;y_{J_2}$ \\
         \\\hline
    \end{tabular}
    \caption{The features used in the kinematic layer of the Multi-layer CNN algorithm.}
    \label{tab:variables}
\end{table}

As shown in Table \ref{tab:3CNN-fixedR}, augmenting the Multi-layer CNN architecture with an additional layer of kinematic features alongside jet images substantially enhances the network performance, initially reaching a significance of 4.2$\sigma$ under ideal conditions and 3.7$\sigma$ with a 10\% systematic uncertainty. These results are further optimised when utilising the variable-$R$ jet clustering algorithm, which pushes the significance to 4.5$\sigma$ in the absence of systematics and maintains a robust 4$\sigma$ even with a 10\% uncertainty in the background estimation (see Table \ref{tab:3CNN-varR}).

\begin{table}[htb!]
    \centering
   \begin{tabular}{|c|c|c|c|}
    \hline
       Process& Cross section$\times$BR (fb)  & Efficiency & Number of events \\\hline
       Signal (S)  & 1.62 & 0.0449 & 21.8\\\hline
       $pp\to bbbb$ (B)  & 1240 & 0.000056  & 20.8 \\ \hline %\hline 
%       S/B &&& 0.174\\\hline
%       $\mathcal{S}$ &&& 1.7  \\\hline
    \end{tabular}
    \vskip 0.3cm 
        \begin{tabular}{|c|c|c|c|}
        \hline
      S & B &  S/B & $\mathcal{S}$ \\\hline
      21.8 & 20.8 & 1.05 & 4.2 \\\hline
    \end{tabular}   
    \caption{Cut efficiency and signal significance after applying the basic acceptance cuts and optimisation of the signal efficiencies using the (double) Multi-layer CNN architecture combined with the kinematic layer at the integrated luminosity $\mathcal{L}=300$ fb$^{-1}$ for the case of the  fixed-$R$ jet clustering algorithm.}
    \label{tab:3CNN-fixedR}
\end{table}

%-------------------------
\begin{table}[htb!]
    \centering
   \begin{tabular}{|c|c|c|c|}
    \hline
       Process& Cross section$\times$BR (fb)  & Efficiency & Number of events \\\hline
       Signal (S)  & 1.62 & 0.0449 & 21.8 \\\hline
       $pp\to bbbb$ (B)  & 1240 & 0.000046 & 17.1\\ \hline %\hline 
%       S/B &&& 0.174\\\hline
%       Significance ($\mathcal{S}$) &&& 1.7  \\\hline
    \end{tabular}
    \vskip 0.3cm 
        \begin{tabular}{|c|c|c|c|}
        \hline
      S & B &  S/B & $\mathcal{S}$ \\\hline
      21.8 & 17.1 & 1.27 & 4.5 \\\hline
    \end{tabular}   
    \caption{Cut efficiency and signal significance after applying the basic acceptance cuts and optimisation of the signal efficiencies using the (double) Multi-layer CNN architecture combined with the kinematic layer at the integrated luminosity $\mathcal{L}=300$ fb$^{-1}$ for the case of the variable-$R$ jet clustering algorithm.}
    \label{tab:3CNN-varR}
\end{table}

\begin{figure}[htb!]
    \centering
    \includegraphics[width=0.75\linewidth]{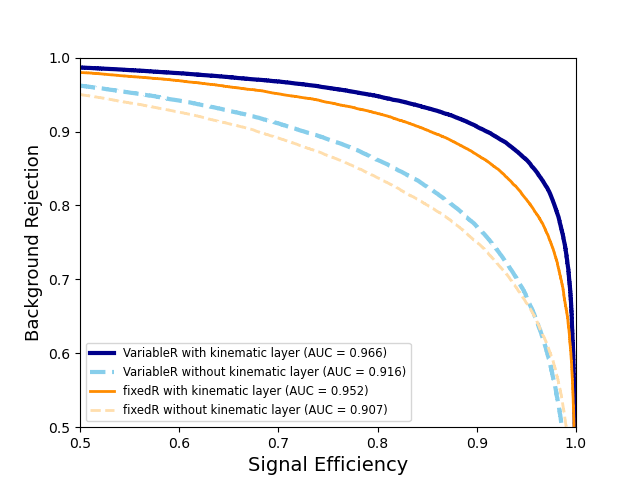}
    \caption{The ROC curves for both fixed- and variable-$R$ jet clustering algorithm are plotted with the $x$-axis representing the signal (extraction) efficiency and the $y$-axis representing the background (rejection) efficiency, with and without an additional kinematic layer.}
    \label{fig:ROC}
\end{figure}

To conclude our analysis, Figure \ref{fig:ROC} illustrates the Receiver Operating Characteristic (ROC) curves for all DL approaches evaluated in this study. The dashed yellow line represents the baseline performance for fixed-$R$ jets without kinematic information, marking the lowest sensitivity. In contrast, the solid blue curve highlights the optimal configuration, achieved by employing variable-$R$ jets integrated with an additional kinematic layer. Other configurations fall in between these two.

% ROC plot 
%--------------------------------------------

%=========================================================

\section{Conclusions} \label{Sec:summary}

To access Higgs self-couplings through di-Higgs production and decay is a priority for Run 3 of the LHC. However, it is  known that establishing a signature therein involving these interactions would imply that they are not of SM origin. In fact, we know that Run 3 of the LHC has no sensitivity to di-Higgs production and decay in the SM. 
In contrast, the latter can occur in BSM scenarios, chiefly, those involving an extended Higgs sector, essentially because one could have SM-like di-Higgs production emerging from the resonant decay of a heavy CP-even Higgs state, $h_2$, with mass larger than $2m_{h_1}=250$ GeV, with $h_1$ being the SM-like Higgs boson discovered at CERN in 2012. Of all such SM-like Higgs boson pairs produced, the vast majority will decay into 4 $b$-quarks, eventually originating boosted jets, if the $h_2$ mass is significantly higher than the $h_1$ one, which has been our assumption throughout. In previous literature (see \cite{Chakraborty:2023hrk}), it was shown that establishing the aforementioned signature could be possible in, e.g., the 2HDM Type-II, via the $gg\to h_2\to h_1h_1$ sub-process,  if the mass of the heavy Higgs state were around 700 GeV or so. In that case, the result was achieved using specific jet clustering algorithms and dedicated kinematic cuts, of course, over limited regions of parameter space of that BSM scenario.

In the present paper, we assess the possibility of doing the same in case of VBF production and promptly established that, using solely the same jet clustering algorithms (either fixed- or variable-$R$ ones) and similar kinematic cuts as those employed for the aforementioned ggF  process, no signature initiated by 4 $b$-quarks, eventually yielding boosted $b$-jets, was accessible. However, in the spirit of Refs.~\cite{Hammad:2023sbd,Hammad:2025wst}, by resorting to a DL analysis exploiting  a Multi-layer CNN architecture deployed on jet images taken in the ($\eta,\phi$) plane after a loose pre-selection exploiting kinematic features separating signal from background, we herein prove that also VBF can indeed be established already at Run 3 of the LHC, using the process $qq\to qqh_2\to qqh_1h_1\to qq$ plus 4 $b$-quarks, in turn yielding 2 forward/backward (slim) jets and 2 or more central boosted $b$-jets. In fact, sensitivity is best when a variable-$R$ algorithm is used instead of a fixed-$R$ one (hence, in line with the findings of \cite{Chakraborty:2023hrk}). We 
show this by using a representative BP from the NMSSM, in the presence of a complete Monte Carlo analysis down to the detector level. 

Hence, as the 4 $b$-quark decay channel of SM-like Higgs pairs is by far the most probable one, we advocate new analyses by ATLAS and CMS based on our findings for the VBF production mode (alongside previous ones  for the case of the ggF case), so as to increase the number of ways in which the self-interactions of Higgs bosons in BSM scenarios can efficiently be tested at the CERN machine.     
%=========================================================
\subsection*{Acknowledgements}
SM is supported in part through the NExT Institute and STFC Consolidated Grant  ST/X000583/1. The authors thank Dr. Shubhani Jain for her inputs in the initial stage of this project. 

\vskip0.75cm
%{\textcolor{green}{Shouldn't the appendix go after the bibliography?}}

%\newpage

%======== References ....====================
%\bibliographystyle{unsrt}
%\bibliography{referance}

\newpage
%\appendix
%\renewcommand{\thesection}{}
\section{Appendix }
\subsection{Architecture and Network Design}
\label{sec:architecture}
 
The analysis employs a multi-layer DL NN architecture specifically designed to exploit complementary information from jet images and high-level kinematic observables (Figure \ref{fig:dnn}). The combined dataset of signal and background events (around 0.5 million samples for each) is partitioned into training, validation and testing samples in the ratio $60:20:20$.
 
For each event, three complementary inputs are processed: the leading and sub-leading jet images, together with a set of discriminating kinematic variables. The two jet images are independently processed through parallel identical image branches while kinematic information flows through a dedicated feature branch based on multi-layer one-dimensional convolutions. The outputs from all three branches are subsequently concatenated and fed into the final classification layers, enabling the network to combine spatial and kinematic information.% in a unified decision boundary.
 
\subsubsection{Image Branches}
\label{subsec:image_branches}
 
The image processing branches, see Figure \ref{fig:MSconv1} (left panel), follow an inception-inspired architecture, a modified version of the traditional CNN architecture, designed to progressively extract hierarchical spatial features from the two-dimensional jet images. Each jet image first passes through an initial convolutional block consisting of three successive convolutional layers with 32, 64 and 128 filters, respectively, each employing $3 \times 3$ kernels followed by batch normalisation, Rectified Linear Unit (ReLU) activation and $3 \times 3$ max-pooling operations. These layers establish a foundation of low- and intermediate level spatial features.
 
The resulting feature maps are then processed through Multi-layer CNN (MLConv1) block, see Figure \ref{fig:MSconv1} (right panel), which are designed to capture information at multiple stages simultaneously. Each MLConv1 block comprises four parallel branches:
\begin{enumerate}
\item[(i)] a $1 \times 1$ convolution producing $f_1$ feature maps;
\item[(ii)] a $1 \times 1$ convolution followed by a $3 \times 3$ convolution producing $f_2$ feature maps;
\item[(iii)] a $1 \times 1$ convolution followed by two sequential $3 \times 3$ convolutions producing $f_3$ feature maps;
\item[(iv)] a $3 \times 3$ max-pooling operation followed by a $1 \times 1$ convolution producing $f_4$ feature maps.
\end{enumerate}
Within each branch, all convolutional layers are accompanied by batch normalisation and ReLU activation. The output of the four branches are concatenated along the channel dimension, enriching the feature representation.% through multi-scale spatial receptivity.
 
The image branch employs two sequential MLConv1 blocks with configuration $(f_1, f_2, f_3, f_4) = (32, 32, 16, 16)$, followed by max-pooling. These are succeeded by two additional MLConv1 blocks with configuration $(64, 64, 32, 32)$. The output of this block is further refined through a convolutional layer with 256 filters, followed by a fully connected layer containing 512 units with ReLU activation. Dropout regularisation with rate of $ 0.2$ is applied both after the MLConv1 blocks and after the dense layer to mitigate overfitting. A global average pooling operation concludes each image branch, producing a compact feature vector of fixed dimensionality. This identical architecture is applied to both the leading and sub-leading jet images, allowing the network to learn event-level features while maintaining symmetry with respect to jet ordering.
 
\subsubsection{Kinematic Branch}
\label{subsec:kinematic_branch}
 
The kinematic branch (see Figure \ref{fig:MSconv2}) is specifically engineered to extract meaningful correlations from high-level observables through a multi-layer one-dimensional convolutional architecture (MLConv2). This design reflects the observation that kinematic variables may exhibit correlations across different scales of abstraction, from pairwise interactions to global event properties. The branch consists of a single MLConv2 block with four parallel pathways of increasing depth:
 
\begin{enumerate}
\item[(i)] a shallow path with a single 1D convolution producing 16 feature maps, followed by global average pooling;
\item[(ii)] an intermediate path with two successive 1D convolutions generating 16 and 32 feature maps, respectively, concluded by global average pooling;
\item[(iii)] a deeper path comprising three convolutional layers (interleaved with max-pooling operations after 2nd and 3rd convolution step), producing 16, 32, and 64 feature maps in sequence;
\item[(iv)] a deep path with four successive 1D convolutional layers producing 16, 32, 64, and 128 feature maps before a final 1D convolution yielding 256 feature maps, followed by global average pooling.
\end{enumerate}
 
All convolutional layers in the kinematic branch incorporate batch normalisation and ReLU activation. The output of all four pathways are concatenated to form a unified feature representation that captures kinematic information across multiple levels of complexity. % This multi-scale structure enables the network to simultaneously exploit both local variable interactions and global correlations within the kinematic feature space.
 
\subsubsection{Classification Layer}
\label{subsec:classifier}
 
The feature vectors derived from the two image branches and the kinematic branch are concatenated and subsequently processed through a series of fully connected layers with 1024, 512, 256 and 128 units, each incorporating ReLU activation. Dropout regularization with rates alternating between $ 0.2$ and $ 0.1$ is applied between the first three successive dense layers to improve generalization performance. The network culminates in a final dense layer with two output nodes and softmax activation, corresponding to the signal and background classification hypotheses. This unified architecture thus provides a principled framework for exploiting the complementary discriminative power of spatial and kinematic information within a single differentiable model.

\end{document}